\documentclass[article]{aa}
\usepackage{graphicx}
\usepackage{graphics}
\usepackage{txfonts}
\usepackage{epsfig}
\usepackage{natbib}
\usepackage{longtable}
\usepackage{lscape}
\usepackage{times}

\begin{document}

\title{A description of sources detected by \emph{INTEGRAL} during the first 4 years of observations}

\author{A.~Bodaghee\inst{1,2}
	\and
	T.J.-L.~Courvoisier\inst{1,2}
	\and
	J.~Rodriguez\inst{3}
	\and
	V.~Beckmann\inst{4}
	\and
	N.~Produit\inst{1,2}
	\and
	D.~Hannikainen\inst{5}
	\and
	E.~Kuulkers\inst{6}
	\and
	D.R.~Willis\inst{1}
	\and
	G.~Wendt\inst{1}}

\authorrunning{Bodaghee~A.~et~al.}
\titlerunning{A description of INTEGRAL sources}
	\offprints{arash.bodaghee@obs.unige.ch}

\institute{
	\emph{INTEGRAL} Science Data Centre, Chemin d'Ecogia 16, CH--1290 Versoix, Switzerland
    \and
	Observatoire Astronomique de l'Universit\'e de Gen\`eve, Chemin des Maillettes 51, CH--1290 Sauverny, Switzerland
    \and
	CEA-Saclay/DSM/DAPNIA/SAp, F-91191 Gif-sur-Yvette, France
    \and
    	NASA Goddard Space Flight Center, Astrophysics Science Division, Greenbelt, MD 20771, USA
    \and
	Observatory, P.O. Box 14, FI-00014 University of Helsinki, Finland
    \and
    	ISOC, ESA/ESAC, Urb. Villafranca del Castillo, P.O. Box 50727, 28080 Madrid, Spain
	}
\date{Received 12-1-07 / Accepted 23-2-07}

\abstract
{In its first 4 years of observing the sky above 20 keV, \emph{INTEGRAL}-ISGRI has detected 500
sources, around half of which are new or unknown at these energies. Follow-up observations at other
wavelengths revealed that some of these sources feature unusually large column densities,
long pulsations, and other interesting characteristics.}
{We investigate where new and previously-known sources detected by ISGRI fit in the parameter space
of high-energy objects, and we use the parameters to test correlations expected from theoretical
predictions. For example, the influence of the local absorbing matter on periodic modulations is
studied for Galactic High-Mass X-ray Binaries (HMXBs) with OB supergiant and Be companions.
We examine the spatial distribution of different types of sources in the Milky Way using various
projections of the Galactic plane, in order to highlight signatures of stellar evolution and to
speculate on the origin of the group of sources whose classifications are still uncertain.}
{Parameters that are available in the literature, such as positions, photoelectric absorption
($N_{\mathrm{H}}$), spin and orbital periods, and distances or redshifts,
were collected for all sources detected by ISGRI. These values and their references are provided
online. }
{ISGRI has detected similar numbers of X-ray Binaries and Active Galactic Nuclei (AGN).
The former group contains new members of the class of HMXBs with supergiant stellar companions.
Usually, this type of object presents strong intrinsic absorption which leads to a peak emission
in an energy range that ISGRI is ideally suited to detect. Thanks to these additional systems, we
are able to show that HMXBs are generally segregated in plots of intrinsic $N_{\mathrm{H}}$ versus
the orbital period of the system and versus the spin period of the pulsar, based on whether the
companion is a Be or an OB supergiant star. We also find a tentative but expected anti-correlation
between $N_{\mathrm{H}}$ and the orbital period, and a possible and unexpected correlation between the
$N_{\mathrm{H}}$ and the spin period.
While only a handful of new Low-Mass X-ray Binaries (LMXBs) have been discovered, there are many sources that remain unclassified and
they appear to follow a spatial distribution typical of Galactic sources (especially LMXBs) rather than
extragalactic sources.
}
{}


\maketitle

\section{Introduction}

In just over 4 years, \emph{INTEGRAL}-IBIS/ISGRI \citep{Leb03,Ube03} has detected
$\sim300$ previously-known sources in the hard X to soft $\gamma$-ray band (20--100 keV), and discovered $\sim200$
sources that were previously unknown at these energies. We will hereafter refer to the latter sources as IGRs
\footnote{an updated list of IGRs can be found at http://isdc.unige.ch/$\sim$rodrigue/html/igrsources.html} (for \emph{INTEGRAL} Gamma-Ray sources). Generally,
these sources were detected by creating long-exposure mosaic images captured
by ISGRI \citep[e.g.,][]{Bir07}. The \emph{INTEGRAL} core programme \citep{Win03} is beginning to
fill in underexposed regions of the sky.

Most of the sources that ISGRI has detected are Low and High-Mass X-ray Binaries (LMXBs and HMXBs, respectively), or Active Galactic
Nuclei (AGN). Both LMXBs and HMXBs feature a compact object such as a neutron star (NS) or a black hole (BH) accreting material from a
companion star: a faint old dwarf in LMXBs (M $\lesssim$ 1 M$_{\odot}$), a bright young giant in HMXBs (M $\gtrsim$ 10 M$_{\odot}$), or
sometimes an intermediate-mass companion. Accretion typically occurs via Roche-lobe overflow in LMXBs or through the wind in HMXBs.
An accretion disk can be found in both types of systems and is an important component of the optical/UV and X-ray emission from AGN and LMXBs.

Subclasses exist within the 3 most common groups. In the case of HMXBs, the spectral type of the stellar companion determines
the sub-classification beyond the NS or BH nature of the compact object. A majority of HMXBs host main-sequence (MS) Be stars that have not
filled their Roche lobe \citep{Wat89}. These systems are usually transient with flares produced whenever the sometimes wide and eccentric orbit
brings the compact object close to its companion. Persistent HMXBs are typically accompanied by an evolved supergiant
(SG) O or B star whose wind steadily feeds the compact object. Their variability stems from inhomegeneities in the wind. Similarly,
LMXBs can be classified based on the type of compact object (NS or BH) it has. Neutron star LMXBs can be divided further into Z or Atoll sources
depending on the tracks they follow in a color-color diagram. The 2 primary groups of AGN are Seyfert
1 and 2, with the latter being more absorbed and showing narrow emission lines only.

\setcounter{table}{1}
\begin{table*}[!t] \centering
	\caption{ The number of sources from each of the major classes
	detected by ISGRI are listed for new ($\equiv$ IGRs) and previously-known sources. Miscellaneous (Misc.)
	sources are Galactic objects that are not X-ray binaries (e.g. CVs, SNRs, AXPs, etc.) and Uncl. refers
	to the group of sources that have yet to be classified. }
	\label{tab_pop}
   	\begin{tabular}{lllllll}	\hline
   	\noalign{\smallskip}
   	\noalign{\smallskip}
       	 			& HMXBs     & LMXBs     & AGN        & Misc.     & Uncl.     & Total  \\
	\noalign{\smallskip}
	\noalign{\smallskip}
	\hline
	\noalign{\smallskip}
	IGRs 			& 32 (15\%) &  6 (3\%)  & 50 (23\%)  & 15 (7\%)  & 111 (52\%) & 214  \\
	previously-known	 	& 46 (16\%) & 76 (27\%) & 113 (40\%) & 32 (11\%) & 18 (6\%)   & 285  \\
	\noalign{\smallskip}
	\hline
	\noalign{\smallskip}
	Total		 	& 78 (16\%) & 82 (16\%) & 163 (33\%) & 47 (9\%) & 129 (26\%)  & 499  \\
	\noalign{\smallskip}
	\hline
   	\end{tabular}
\end{table*}

Our understanding of the different populations of \emph{INTEGRAL} sources is limited by the large number of sources about which very little is known.
Roughly half of all IGRs remain unclassified. The nature of these sources is difficult to elucidate given that many are faint or transient.
Furthermore, the images, spectrum, and timing analysis gathered from a single energy range are usually insufficient to classify an object. Information from other
wavelengths such as soft X-rays, infrared or radio are necessary to help identify the nature of a source. For example, radio emission can be the signature of a
jet or pulsar, while the optical spectral type can help distinguish between LMXBs and HMXBs, and the redshift can place it at extragalactic distances.
Follow-up observations with soft X-ray telescopes (i.e. \emph{Chandra}, \emph{RXTE}, \emph{Suzaku}, \emph{Swift} and \emph{XMM-Newton}) can
provide fine timing analyses which enable short-period modulations to be found, and they can describe the shape of the continuum below
ISGRI's $\sim$20 keV lower limit, in an energy range where potential photoelectric absorption
($N_{\mathrm{H}}$) and iron fluorescence lines are detectable. Precise X-ray coordinates from \emph{Chandra}, \emph{Swift} or \emph{XMM-Newton} can be used to
search for counterparts in dedicated radio, optical, and IR observations or in catalogues. However, many sources are clustered in the
Galactic center and along the plane, which, because of the density of stars and the amount of obscuring dust, can hinder the identification of
the optical/IR counterpart.

Perhaps the most interesting result from follow-up observations is that a number of IGRs present column densities that are much higher than would be
expected along the line of sight. These large absorptions are therefore intrinsic and could be the reason these sources eluded discovery with previous (softer)
X-ray missions. The first new source discovered by \emph{INTEGRAL} is IGR J16318$-$4848 \citep{Cou03} which is one of the most absorbed Galactic sources known with
$N_{\mathrm{H}}\sim 2\cdot 10^{24}$ cm$^{-2}$ or roughly 2 orders of magnitude more than the intervening Galactic material \citep{Dic90}. Since this
discovery, other sources joined the growing class of heavily-obscured X-ray sources described by \citet{Wal04} and \citet{Kuu05}.

A certain number of these absorbed sources consist of X-ray pulsars: e.g. IGR J16320$-$4751 \citep{Rod06},
IGR J16393$-$4643 \citep{Bod06}, and IGR J17252$-$3616 \citep{Zur06}. Their persistent emission,
their long pulse periods ($\sim 1$ ks), and their large column densities suggest that these systems are likely to be SG HMXBs, with the NS deeply
embedded in the wind of its massive stellar companion \citep{Wal06}. These kinds of systems are still a minority compared to Be HMXBs,
but \emph{INTEGRAL} is expanding their ranks.

Supergiant Fast X-ray Transients (SFXT: see \citet{Sgu05} for a review) are another type of object whose numbers are increasing thanks
to \emph{INTEGRAL}. These objects are HMXBs whose X-ray emission is characterised by short strong outbursts (a peak flux of up to a Crab or more during a
few seconds to a few hundred seconds), sometimes with recurrence timescales that can reach several hundred days. Despite the intensity of their outbursts,
SFXTs are usually not detected in deep mosaic images because the accumulation of exposure time attenuates their significance. Therefore, the search for
SFXTs involves scanning archival light curve data and short-exposure mosaic images for rapid bursts from known transients or at new locations.

ISGRI has also detected other types of Galactic objects such as Cataclysmic
Variables (CVs), Supernova Remnants (SNRs), Pulsar Wind Nebulae (PWN), Anomalous X-ray Pulsars (AXPs), etc.,
which are referred to henceforth as Miscellaneous.

The most extensive catalogues of sources detected by ISGRI are the catalogues of \citet{Bir06} and \citet{Bir07}.
These catalogues represent fairly large and homogeneous samples that can be used to
study the general characteristics of populations of high-energy sources \citep[e.g.][]{Leb04,Dea05,Lut05,Bec06b}.

This research presents the parameters of all sources detected by ISGRI and reported between
its launch on Oct. 17, 2002, until Dec. 1, 2006. Absorption values, pulse and
orbital periods, and distances or redshifts were
collected from the literature and were used to study the populations of high-energy sources,
to test various correlations expected from theoretical
predictions, and to investigate where the new and previously-known sources detected by ISGRI
fit in the parameter space of high-energy objects.

\begin{figure*}[!t] \centering
\includegraphics[width=15cm,angle=0]{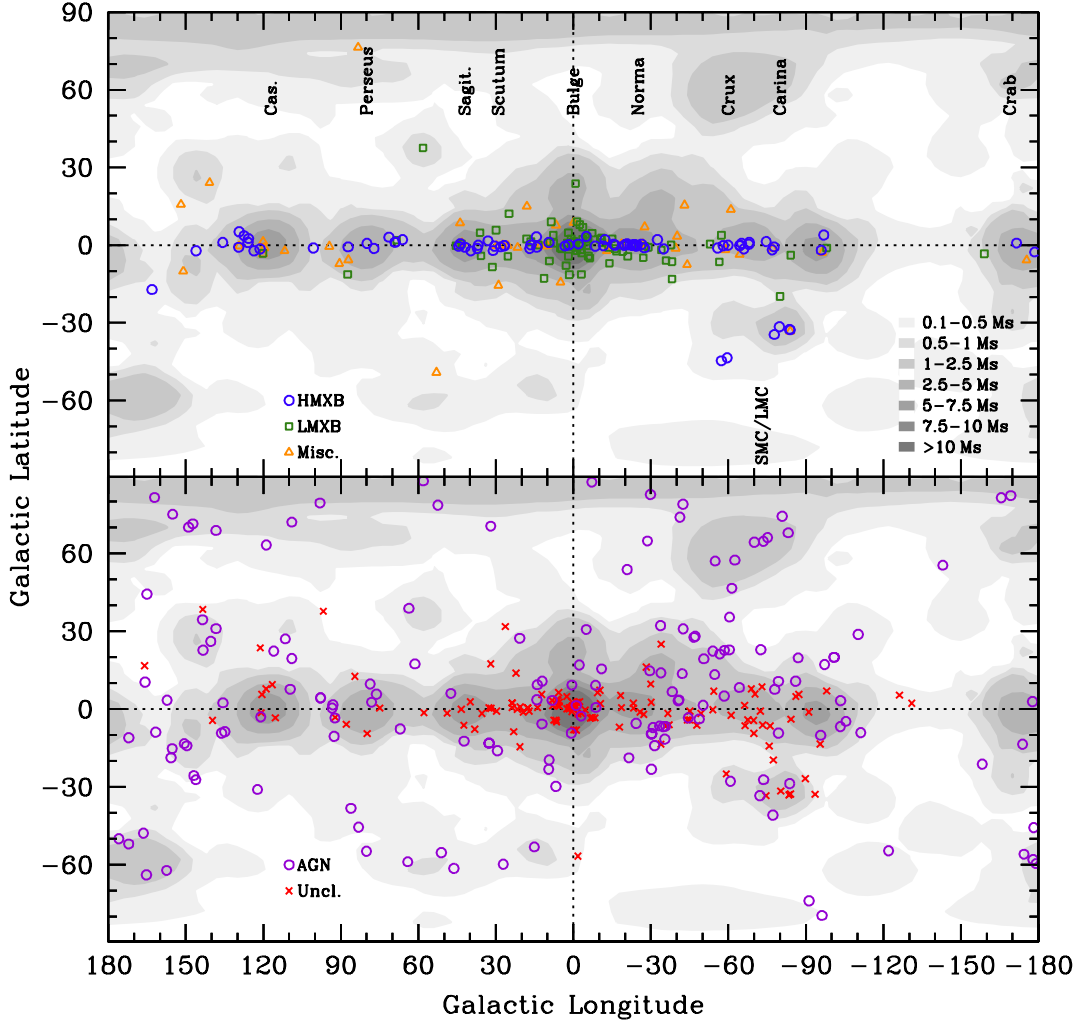}
\caption{Spatial distribution in Galactic coordinates of sources detected so far by ISGRI.
The figure at the top presents the distributions of HMXBs (stars), LMXBs (squares) and
miscellaneous sources (triangles). The figure at the bottom
displays extragalactic sources (circles) and unclassified sources (crosses).
The directions to the spiral arm tangents and
other areas of interest are indicated, as are the cumulative exposure times at each location (from public data
in revs. 30--484). The number of sources in each class is listed in Table\,\ref{tab_pop}.}
\label{fov_disp}
\end{figure*}

\section{Data \& Analysis}
\label{data}

We selected all sources from Version 27 of the \emph{INTEGRAL} General Reference Catalogue \citep{Ebi03} which were detected by ISGRI
(i.e. those with ``ISGRI\_FLAG==1''). These flags were set to the value of 1 as soon as confirmation of a detection by ISGRI is announced in an article,
conference proceeding, Astronomer's Telegram or IAU circular. Therefore, the completeness of the sample is difficult to evaluate given that
by definition, the present sample contains all sources that were detected above 20 keV while within the ISGRI FOV at some point during the last
$\sim$4 years, without considering the detection significance, nor the amount of exposure time that was required to make the detection.

The exposure map that can be seen in Fig.\,\ref{fov_disp} was created by accumulating all public pointings
in revolutions 30--484 (UTC: 11/1/2003--1/10/2006). Due to the core programme
observation strategy, the Galactic centre (GC) is heavily exposed ($t_{exp}>10$ Ms) whereas some regions have less than 10 ks of exposure time dedicated
to them. The exposure is uneven along the Galactic plane as well, with exposure biases in the
directions of the spiral arms. The sensitivity limit of a source in the most exposed regions is as low
as $\sim1$ mCrab for a transient object detected at the 6$\sigma$ level \citep{Bir06}. In fact, some sources were
detected only because the instrument serendipitously caught a flaring event. Using the fact that the Log($N$)--Log($S$)
relation for extragalatic sources follows a power law with a slope of $-3/2$ \citep{For78}, we can estimate a
sensitivity limit of $\lesssim5$ mCrab ($=3.78 \cdot 10^{-11}$ ergs cm$^{-2}$ s$^{-1}$ in 20--40 keV)
for our sample based on where the distribution of AGN from \citet{Bir06} deviates from the expected slope.

A number of IGRs have soft X-ray counterparts that were sometimes detected by earlier missions.
For example, IGR J16393$-$4643 was known as AX J1639.0$-$4642
by \emph{ASCA}, and IGR J17252$-$3616 as EXO 1722$-$360 by \emph{EXOSAT}, and many IGRs have ROSAT counterparts \citep{Ste06}. Since
ISGRI was the first to detect them above 20 keV, it is legitimate to group them together as a population of new soft $\gamma$-ray sources.
They can then be compared to sources detected by ISGRI that were previously known to emit above 20 keV (e.g., Crab, Vela X-1, etc.).
Note that the so-called previously-known sources actually include a few objects that were discovered after the launch of \emph{INTEGRAL}
(e.g. by \emph{HETE}, \emph{RXTE} or \emph{Swift}).

\begin{figure*}[!t] \centering
\includegraphics[width=12cm,angle=0]{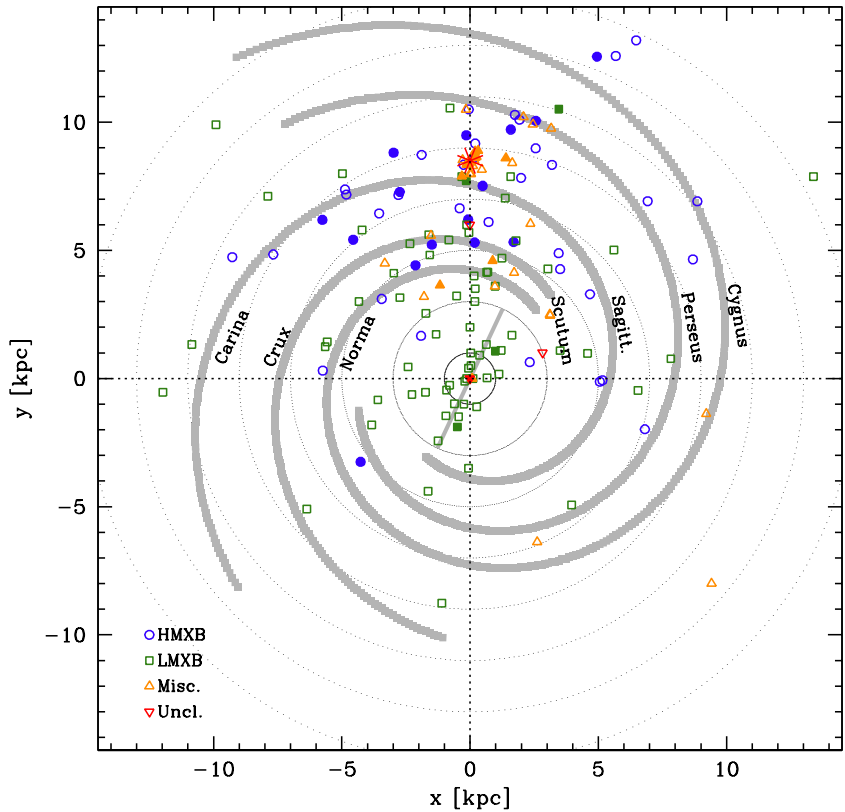}
\caption{Galactic distribution of HMXBs (49, circles), LMXBs (74, squares),
miscellaneous sources (37, triangles), and unclassified sources (3, inverted triangles).
Filled symbols represent IGR sources. Also plotted is the 4-arm Galactic spiral model from \citet{Rus03}
with the Sun located at 8.5 kpc from the centre. The
concentric circles indicate radii of 1, 3, 5,..., kpc from the centre.}
\label{gal_disp}
\end{figure*}

The name or position of each source in our sample was queried to the SIMBAD and ADS servers for references that could provide any
of the following parameters: position and error radius, classification, column density ($N_{\mathrm{H}}$), spin period, orbital period,
and distance (or redshift). Besides a rough X-ray position, very little is known about some sources, while other sources were
so thoroughly studied that choices had to be made between sometimes conflicting values (notably $N_{\mathrm{H}}$ and distance).
The index of parameters that we have constructed (see Table\,1 \footnote{Table\,1 is only available in
electronic form at the CDS via anonymous ftp to cdsarc.u-strasbg.fr (130.79.125.5) or via
http://cdsweb.u-strasbg.fr/Abstract.html}) represents what we know about sources detected by ISGRI to date
(until December 1, 2006). The structure of Table\,1 is as follows:

\begin{itemize}
\item {\em Name}:  \\
Most sources have more than 1 name owing to detections by various instruments operating at different energies.
As in \citet{Ebi03}, we selected names that are commonly used in high-energy astrophysics and that
are accepted as an identifier in SIMBAD. This eases comparisons with other catalogs.
\item {\em Position}: \\
Source positions are from the X-rays unless a more accurate position at other wavelengths is known
for a confirmed counterpart. Right Ascension and Declination in J2000 coordinates are given in
``hh mm ss.s'' and ``deg arcmin arcsec'' formats, respectively. The uncertainty (in arcmin) of the position from
the reference is given as an individual entry, and it is reflected in the representation
of the source positions. Galactic coordinates are also provided.
\item {\em Absorption}: \\
Column densities (in $10^{22}$ cm$^{-2}$) were gathered from the literature whenever a model fit
to the X-ray spectrum required an absorption component. Extracting a single $N_{\mathrm{H}}$ for a source and
comparing this value to those of other sources is not a
straightforward exercise since intrinsic column densities are not static. A measurement made
during flaring or quiescent periods, or at different orbital epochs, will
heavily influence the $N_{\mathrm{H}}$. The geometry of the system, the energy range of the
satellite that gathered the data, and the model used to describe the resulting spectrum also affect the
$N_{\mathrm{H}}$ value. Therefore, the uncertainties are often large or only upper limits are provided.
Whenever possible, we selected the $N_{\mathrm{H}}$ value of the model that best fits a recent X-ray spectrum
taken with a telescope that covers the soft X-ray domain well.
\item {\em Modulations}: \\
Spin periods (in seconds) and/or orbital periods (in days) have been reported for a large number of
Galactic objects detected by ISGRI. Some systems are known to spin down, so we selected
the most recent value from \emph{RXTE} whenever possible, even though this level of precision is not needed for
the purpose of our study. The spin period can refer to the spin of the NS in X-ray binaries, or to
the spin of the White Dwarf in CVs. The catalogs of \citet{Liu00} and \citet{Liu01}, and the
systematic analysis of \emph{RXTE} data by \citet{Wen06} are among the main references of spin and orbital
periods in this work.
\item {\em Distances}: \\
The distances to extragalactic objects are given as a measure of the redshift (denoted by brackets).
Objects in the Milky Way and Magellanic Clouds have distances in units of kpc.
Opinions sometimes differ as to the distances of some Galactic sources. We favored distance
measurements that were recent, precise and that were the least model-dependent. Even so, the distance
uncertainties quoted in the literature can be large. The distance measurements are from, among others,
\citet{Whi96}, \citet{Gri02}, \citet{Jon04}, \citet{Bas06}, \citet{Bec06a}, or
from references in \citet{Liu00} and \citet{Liu01}. Note that the distances in
\citet{Gri02} sometimes represent an average over several competing distance estimates.
\item {\em Type}: \\
Source classifications are based on the concensus opinion in the literature.
Peculiar behaviour, such as quasi-periodic oscillations, transience, and Z-track or Atoll shapes, are also noted
since they can help us distinguish between systems within the same class. If the source classification has
not yet been confirmed, it is simply called ``unclassified.'' The transient identification
of a source is given by any of the following: 1) the label was assigned by its
discoverers or by other authors (e.g. "Discovery of a new \emph{transient} IGR J...");
2) the source has not been detected by anyone else since its discovery announcement, e.g.
it is not listed in the all-data, all-sky catalogs of \citet{Bir06} and \citet{Bir07});
3) the source is detected only in mosaic images of a single or a few
consecutive revolutions according to \citet{Bir07}, but not in their all-data
mosaic images.
\item {\em References}: \\
All parameters are referenced so that the reader can assess the methods that were used to determine the
quoted values and uncertainties.
\end{itemize}

\begin{figure}[!t] \centering
\includegraphics[width=8cm,angle=0]{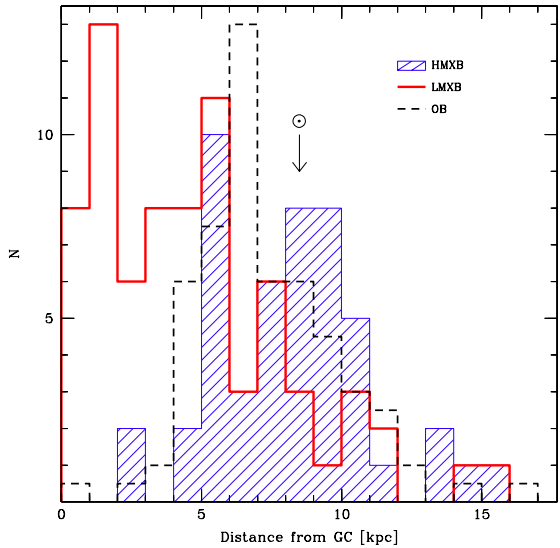}
\caption{Distribution of galactocentric distances of HMXBs (49, shaded histogram) and
LMXBs (74, thick histogram). The dashed histogram represents OB star-forming complexes
from \citet{Rus03} (divided by 2).}
\label{hist_dist}
\end{figure}

\section{Results}
\label{res}

\subsection{Spatial Distribution}

Table\,\ref{tab_pop} lists the major source populations detected by ISGRI that are either new
($\equiv$ IGRs) or that were previously known. ISGRI has discovered many new HMXBs and AGN, but their proportions
relative to the other classes are similar to what was known before the launch of \emph{INTEGRAL}. Only a few
LMXBs have been discovered by ISGRI. This is because LMXBs are generally less intrinsically obscured
than HMXBs or AGN, and are therefore easier to detect with previous satellites. Around 50 sources have been detected
that belong to the group of miscellaneous sources (i.e. CVs, SNRs, PWN, AXPs, etc.), while $\sim130$ IGRs await
classification.

\begin{figure}[!t] \centering
\includegraphics[width=8cm,angle=0]{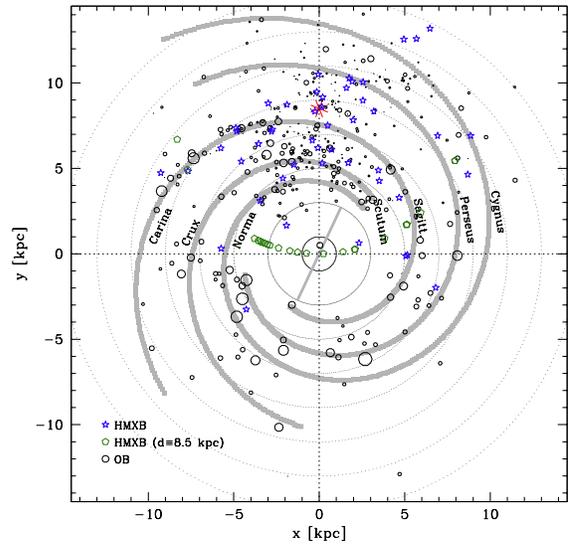}
\caption{Galactic distribution of HMXBs whose distance are known (49, star symbols)
and the locations of star-forming complexes from \cite{Rus03} (464, circles).
The symbol size of the latter is proportional to the activity of the complex. HMXBs whose
distances are unknown have been placed at 8.5 kpc (23, pentagons).}
\label{gal_disp_hmxb}
\end{figure}

\begin{figure}[!t] \centering
\includegraphics[width=8cm,angle=0]{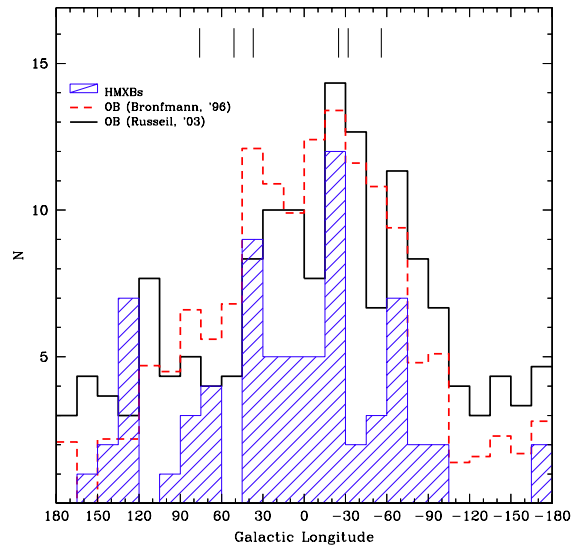}
\caption{Histograms of Galactic longitudes integrated over the
latitude for HMXBs with $|b|<6^{\circ}$ (shaded histogram)
and star-forming regions from \citet{Rus03} (thick histogram, divided by 3)
and from \citet{Bro96} (dashed histogram, divided by 10). The vertical lines indicate
the tangential directions of the 4-arm spiral model from \citet{Rus03}}
\label{fov_disp_hmxb_OB}
\end{figure}

The spatial distributions, in Galactic coordinates, of the major classes of $\gamma$-ray sources detected by ISGRI are presented in Fig.\,\ref{fov_disp}.
Naturally, ISGRI detects sources in regions that are exposed. Given the heterogeneous exposure map of the sky
gathered in the last 4 years of observations, detections are biased towards regions of the sky that have
been exposed the longest (i.e. the Galactic plane and bulge).

However, Fig.\,\ref{fov_disp} also demonstrates the effect that the evolution
of each type of source has on its spatial distribution. Because their optical companions belong to an old stellar population, LMXBs are found
predominantly in the Galactic bulge and/or they have had time to migrate off the plane of the Milky Way ($|b| \gtrsim 3$--$5^{\circ}$).
On the other hand, the stellar companions of HMXBs are young stars, so these systems must remain close to sites of recent stellar formation. Thus,
the angular distribution of HMXBs reflects the spiral structure of the Galaxy, with an uneven distribution along the Galactic plane punctuated by
peaks that are roughly consistent with the tangential directions to the inner spiral arms. Those HMXBs that have been detected at longitudes
$|l| \gtrsim 90^{\circ}$ correspond to systems located around spiral arms near the Sun. Evolutionary signatures like these were noticed in the past by
\emph{Ginga} \citep{Koy90}, \emph{RXTE} \citep{Gri02}, and more recently with \emph{INTEGRAL} \citep{Dea05,Lut05}, although their samples were smaller
than the one presented here.

\begin{figure}[!t] \centering
\includegraphics[width=8cm,angle=0]{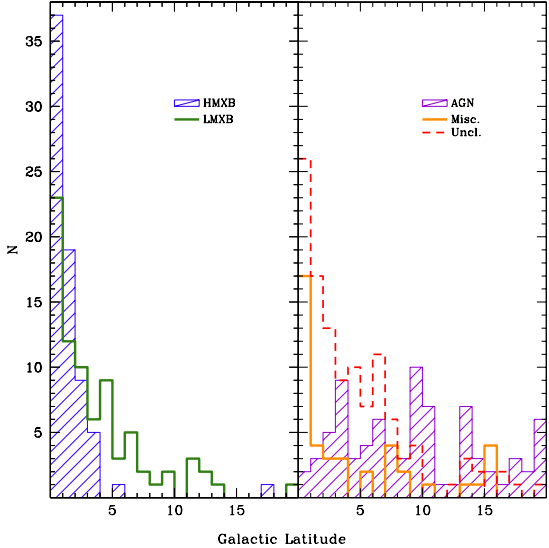}
\caption{Angular distribution (in degrees, for $|b| < 20^{\circ}$) from the Galactic plane of
sources that have been detected by ISGRI. Shaded histograms are HMXBs (left) and AGN (right), and
the clear histograms represent LMXBs (left) and miscellaneous sources (right). The dashed
histogram denotes unclassified sources. The distributions have been summed over the northern and
southern Galactic hemispheres. The curves represent fits to the data from the model
and parameters described in the text.}
\label{hist_lat}
\end{figure}

\begin{figure}[!t] \centering
\includegraphics[width=8cm,angle=0]{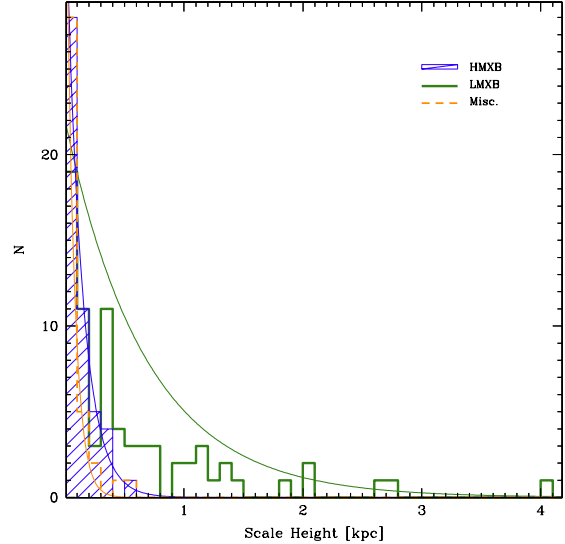}
\caption{\small{Vertical scale height (in kpc) from the Galactic plane of
HMXBs, LMXBs and miscellaneous sources whose distances are known.
The distributions have been summed over the northern and southern
Galactic hemispheres. Sources from the Magellanic Clouds are excluded. The curves
represent the exponential model described in the text fit to the data
(see Table\,\ref{tab_height} for parameters). }}
\label{hist_height}
\end{figure}

Another way to demonstrate the role of stellar evolution in shaping the spatial distributions of
LMXBs and HMXBs is to plot the positions of sources whose distances are known on a spiral arm
model of the Milky Way. \citet{Rus03} developed the Galactic spiral arm model that we used.
Their model is based on the  locations of star-forming complexes that include groups of OB stars,
molecular clouds, H\,II regions, and diffuse ionised gas. The locations of these complexes are
derived from a variety of tracers such as H$\alpha$, CO, the radio continuum and absorption lines.

While the uncertainties on distances can be large, Fig.\,\ref{gal_disp} shows that HMXBs tend to occupy the outer
disk and arms where young stars are formed, whereas LMXBs are clustered near the bulge where old globular
clusters reside. A histogram of galactocentric radii (Fig.\,\ref{hist_dist}) shows LMXBs peaked at the center and
decreasing gradually, while HMXBs roughly follow the distributions from H\,II/CO surveys \citep{Rus03}
which are underabundant in the central few kpc and peak at the spiral arms. According to a
Kolmogorov-Smirnov (KS) test, the probability is less than 0.01\%
that the galactocentric distributions of LMXBs and HMXBs are statistically compatible.

\begin{table}[!t] \centering
         \caption{ Parameters from the model described in the text fit
	to the distributions of scale heights from the Galactic plane for
	HMXBs, LMXBs and miscellaneous sources whose distances are known.
	Objects from the Magellanic Clouds are excluded.}
         \label{tab_height}
         \begin{tabular}{lccc}        \hline
         \noalign{\smallskip}
         \noalign{\smallskip}
	& $k$ & $\alpha$ & $h_{0}$ [pc] \\
         \noalign{\smallskip}
         \noalign{\smallskip}
         \hline
         \noalign{\smallskip}
         HMXBs & $36\pm3$ & $7.5\pm1.7$ & $134_{-25}^{+39}$   \\
         \noalign{\smallskip}
         LMXBs & $22\pm3$ & $1.5\pm0.5$ & $680_{-160}^{+320}$   \\
         \noalign{\smallskip}
         Misc. & $41\pm3$ & $15\pm4$ & $66_{-14}^{+23}$   \\
         \noalign{\smallskip}
         \hline
         \end{tabular}
\end{table}

The distribution of LMXBs in the central Galaxy suggests an association with the
Galactic bar (Fig.\,\ref{gal_disp}). Low-mass X-ray binaries whose distances are known
and that have been detected by ISGRI are not prevalent on either side of the bar.
Only 1 LMXB with a distance measurement has been detected in the Galactic center region
bound by $0 < x < 3$ kpc and $-3 < y < 0$ kpc, indicating that the bar might be responsible
for preventing an identification and distance measurement to be made for the faint
counterparts to LMXBs situated behind it. As viewed from the Sun, the orientation of the bar leads
to an apparent asymmetry of LMXBs in the central 3 kpc of the Galaxy ($|l| \lesssim 20^{\circ}$):
in this direction, ISGRI has detected 50\%
more LMXBs at negative longitudes than at positive longitudes. Maps of
Galactic absorption are expected to be symmetrical in this region \citep{Dic90}.

In Fig.\,\ref{gal_disp_hmxb}, we present the Galactic distribution of star-forming
complexes \citet{Rus03} with the symbol size proportional to the
excitation parameter in that region ($\equiv$ amount of ionising photons as determined from the
radio continuum flux). High-mass X-ray binaries whose distances are known are
symbolised by stars, while those with unknown distances were assigned a distance of 8.5 kpc and
are represented by pentagons. The 4-arm spiral model of \citet{Rus03} is also drawn.
Figure\,\ref{fov_disp_hmxb_OB} presents histograms of
Galactic longitudes (integrated over the latitude) of HMXBs (shaded histogram, with $|b|<6^{\circ}$,
in order to exclude sources in the Magellanic Clouds). Also shown are angular distributions of
star-forming complexes from \citet{Rus03} (divided by 3, thick histogram), and of
ultra-compact H\,II regions detected by \emph{IRAS} \citep{Bro96} (divided by 10, dashed histogram).

In general, the distribution of HMXBs along the
plane of the  Milky Way coincides with the expected radial distribution of young massive star-forming
regions. A KS test yields a probability of 22\%
that the distributions of HMXBs and \emph{IRAS} sources
(shaded and dashed histograms, respectively, in Fig.\,\ref{fov_disp_hmxb_OB}) are
statistically compatible. Excluding HMXBs that lie outside the survey region covered by \citet{Bro96}
($|b|<2^{\circ}$ for $|l|<60^{\circ}$, and $|b|<4^{\circ}$
elsewhere) increases the probability of statistical compatibility to 34\%.
Peaks at Galactic longitudes $l\sim\pm30^{\circ}$ are observed in both data sets corresponding to the
direction of the inner spiral arm tangents (Norma and Scutum/Sagittarius arms). \citet{Bro96}
remark that the peaks in their distribution are also consistent with another active formation site of
young, massive stars: a molecular ring situated at a radius of $\sim$3 kpc from the Galactic center.

At first glance, the distributions of star-forming complexes of \citet{Rus03} and HMXBs
are also compatible (thick histogram in Fig.\,\ref{fov_disp_hmxb_OB}).
The KS test returns a probability of only 3\% which is misleading given the large number of
objects in \citet{Rus03} that are not very active. When selecting complexes with an excitation
parameter $>$ 10 pc cm$^{-2}$, which still represents 70\%
of the sample, the statistical compatibility improves to 41\%.

\citet{Lut05} and \citet{Dea05} found that the distribution of HMXBs was offset with respect to the
directions of the spiral arm tangents. \citet{Lut05} note that $\sim10$ Myr must elapse before one
of the stars in a binary system collapses into a NS or BH, and that Galactic rotation will induce
changes in the apparent positions of the arms relative to the Sun. This implies a delay between the
epoch of star formation and the time when the number of HMXBs reaches its maximum.
The observed displacement could simply stem from uncertainties in the distances to the HMXBs.
Another problem is that the exact location of the arms depends on which Galactic model is used.
Changes in the Sun-GC distance or in the pitch angles of the arms affects the radial scaling and
shifts the tangential directions.

The propagation of density waves is believed to promote star formation in the spiral arms
\citep{Lin69}. Depending on the distance to the GC, the spiral arm pattern has
angular velocities in the range of $\Omega\sim$20--60 Gyr$^{-1}$ \citep{Bis03}. Hence, in the last
$\sim10$ Myr (corresponding roughly to the epoque when the inner spiral arms and the current density
maxima of HMXBs overlapped), the inner arms of the Galaxy have rotated around the GC by
$\sim40^{\circ}$. Individual stars (such as the Sun) or groups of stars have
negligible movement in this scenario. Instead, star-forming sites that are now active
(e.g. from \citet{Rus03}) should be about $\sim40^{\circ}$ away from those regions that were active
some 10 Myr ago and that produced the current crop of HMXBs. Therefore, in order to reproduce the
distribution of active star-forming sites as they were some 10 Myr ago, we
introduced differential Galactic rotation to ``unwind'' the distribution of
current star-forming complexes from \citet{Rus03}. Kolmogorov-Smirnov tests suggest that the
effects of Galactic rotation are negligible, even when only the most active sites are considered.

The distribution of angular distances from the Galactic plane (in degrees, for $|b| < 20^{\circ}$)
of sources detected by ISGRI is shown in Fig.\,\ref{hist_lat}. The distributions were summed over the
northern and southern Galactic hemispheres. Shaded histograms are HMXBs (left) and AGN (right), and the
thick histograms represent LMXBs (left) and miscellaneous sources (right). The distribution of
unclassified sources is given by the dashed histogram. Not surprisingly, the spread of the latitude
distributions is larger in LMXBs than it is in HMXBs owing to the relative youth of the optical
primaries in the latter. Also expected is the distribution of AGN which is more or less flat and which
roughly follows the exposure map. However, the Galactic plane ($|b|\lesssim 3^{\circ}$) is noticeably
deficient in AGN detections despite the fact that the exposure map is biased here. This highlights
the difficulty in detecting AGN at high energies and at low latitudes; these objects tend to be
intrinsically absorbed, they are further obscured by the Galactic plane, and their counterparts have to
be identified within a crowded field. \citet{Saz07} noted that the exclusion of sources in the Galactic
plane region ($|b|<5^{\circ}$) from an all-sky survey resulted in only a marginal reduction in the
number of identified AGN, whereas the number of unclassified sources dropped significantly.

It is useful to examine the unclassified sources as they help to define the limits of our study. Almost all of
the sources that are unclassified have position accuracies that are no better than a few arcminutes. This
precludes establishing an optical counterpart for many unclassified sources located in crowded regions such as
the Galactic plane. The transient nature of many unclassified sources implies a lack of
immediate follow-up observations that would permit a classification. Because of their transience, many
unclassified sources appear fainter than average in long-exposure mosaic images.
\citet{Mas06} suggest that up to half of all unclassified sources could be AGN situated behind
the Galactic plane, whereas \citet{Dea05}, working on a smaller sample, favor a Galactic origin for the
unclassified sources of \citet{Bir04} based on the slope of the Log($N$)--Log($S$) relation and other factors.

Unclassified sources have a distribution of Galactic latitudes that peaks in the central 3$^{\circ}$ from the
Galactic plane and decreases gradually, suggesting a population of sources that are Galactic rather
than extragalactic in origin (see Fig.\,\ref{hist_lat}). Many of the unclassified sources also happen to be
transient, whereas AGN can vary but tend to emit persistently. Furthermore, the
angular distribution of unclassified sources is very similar to the distribution of LMXBs with a KS-test
probability of nearly 40\%
of statistical compatibility between unclassifieds and LMXBs, compared with 13\%
for miscellaneous sources, and less than 0.01\%
for either AGN or HMXBs.

While there are extragalactic sources among them, the population of unclassified sources is
therefore likely to be composed primarily of Galactic sources such as
LMXBs and miscellaneous sources. We are unable to elaborate on the proportions of the different
classes, but it is clear from Fig.\,\ref{hist_lat}, and from the results of KS tests, that the
unclassified sources are most similar to the LMXBs and miscellaneous sources in
their distribution off the Galactic plane. The reasons they have avoided classification (and
detection by previous missions) are: the companion stars to LMXBs are usually faint in the optical/IR spectrum;
they are located near the Galactic plane where absorption and source confusion prevent an identification; and,
for many of these sources, their transient emission complicates efforts to perform follow-up observations.
Recent improvements in Target of Opportunity campaigns aimed at new IGRs have uncovered as many new
LMXBs in the last year than during the first 3 years of observations combined.

\begin{figure}[!t] \centering
\includegraphics[width=8cm,angle=0]{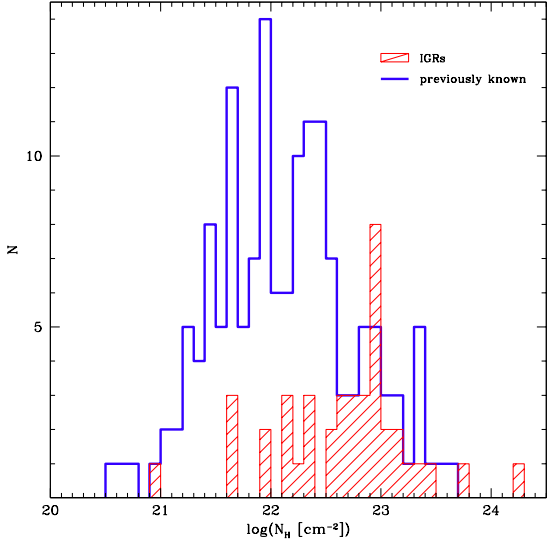}
\caption{The distribution of reported column densities ($N_{\mathrm{H}}$) for
Galactic sources (including sources in the Magellanic Clouds) detected by ISGRI that were previously known
(152, clear histogram) and for IGRs (41, shaded histogram).}
\label{hist_nh}
\end{figure}

Figure\,\ref{hist_height} presents the distributions of scale heights (in kpc) for
HMXBs (shaded histogram) and LMXBs (clear histogram) whose distances are known. Sources from
the Magellanic Clouds are excluded.
Following the procedure in \citet{Dea05}, we set the number of sources as a function of the
distance in kpc above the Galactic plane ($h$) according to $N = k \cdot e^{-\alpha\cdot h}$ where
$\alpha \equiv 1/h_{0}$ describes the steepness of the exponential.
The parameters that best fit this model are listed in Table\,\ref{tab_height}. The value that we
derive for the characteristic scale height ($h_{0}$) of HMXBs is $\sim130$ pc which is compatible with the
value found by \citet{Gri02} with \emph{RXTE} data, but slightly less than the value
from \citet{Dea05} ($\gtrsim200$ pc). The characteristic
scale height that we derive for LMXBs ($\sim600$ pc) is larger than the scale heights found
by \citet{Gri02} and \citet{Dea05} which were closer to $\sim400$ pc. This is probably due to the greater coverage of the sky and larger sample size of our
study. Miscellaneous sources have a distribution that is more similar to HMXBs than it is to LMXBs.

\begin{figure*}[!t] \centering
\includegraphics[width=15cm,angle=0]{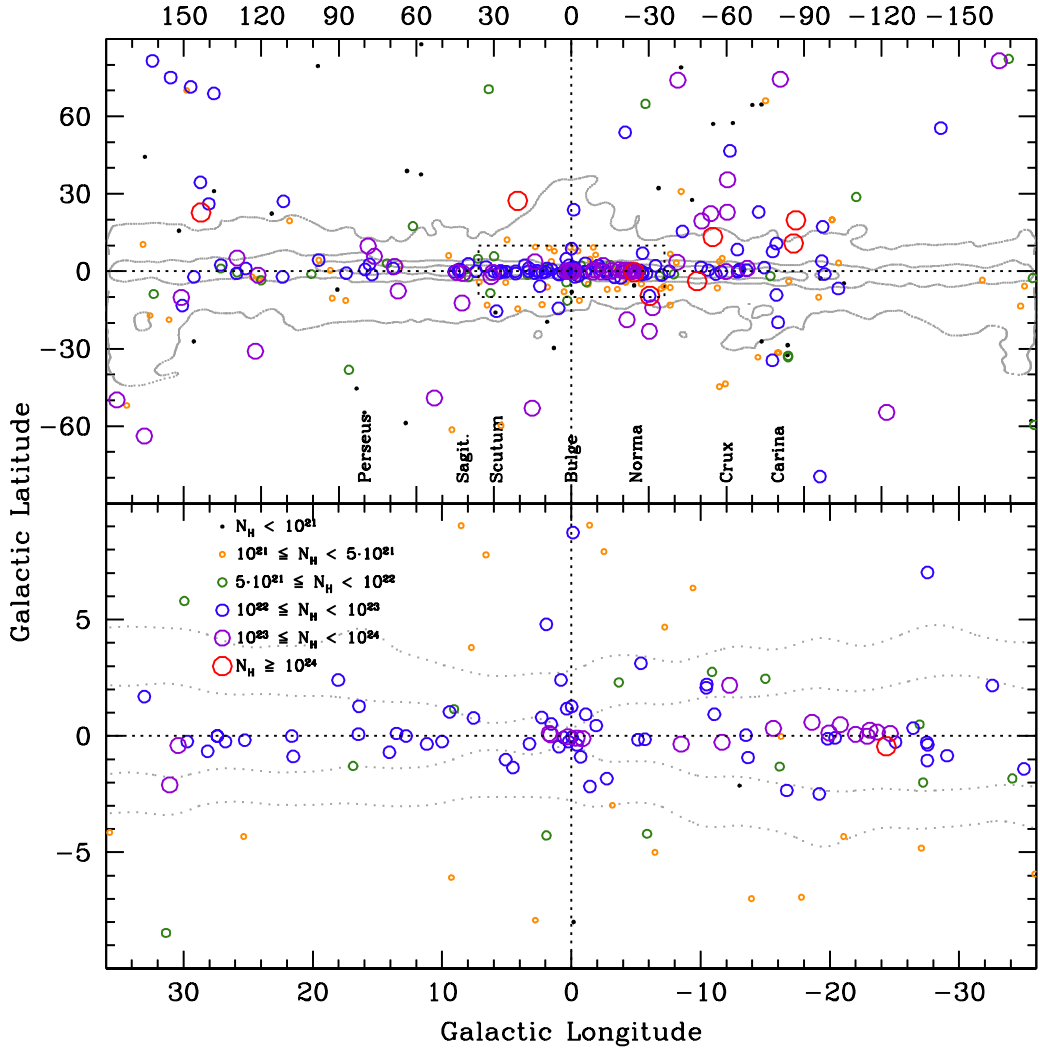}
\caption{Spatial distribution, in Galactic coordinates, of all sources detected
by ISGRI for which $N_{\mathrm{H}}$ has been reported. The symbol size is proportional to the
published column density. The figure at the top shows the whole sky and includes extragalactic
sources, while the figure at the bottom focuses on the Bulge region
(boxed region in the figure at the top) and excludes extragalactic sources.
Contours denote Galactic absorption levels of
$10^{21}$, $5\cdot 10^{21}$, and $10^{22}$\,cm$^{-2}$ \citep{Dic90}.}
\label{fov_disp_nh}
\end{figure*}

\subsection{Absorption}

The column densities along the light of sight of some sources in our sample are higher than the value
expected from radio maps \citep{Dic90} which implies absorbing material intrinsic to the source.
On average, Galactic IGRs are more absorbed than the sources seen before \emph{INTEGRAL} (by a factor of $\sim$4) with
IGRs representing a sizable contingent of objects that have
$N_{\mathrm{H}}\sim 10^{23}$ cm$^{-2}$ (see Fig.\,\ref{hist_nh}). The average column density of sources that were previously known is
$N_{\mathrm{H}}=1.2 \cdot 10^{22}$ cm$^{-2}$ ($\sigma \sim 0.7$) whereas IGRs have an average
$N_{\mathrm{H}}=4.8 \cdot 10^{22}$ cm$^{-2}$ ($\sigma \sim 0.6$). A KS test
yields a probability of less than 0.01\%
that the two distributions are statistically compatible.

The classified Galactic IGRs are mostly HMXBs (Table\,\ref{tab_pop}) which usually exhibit high column densities, either
intrinsically due to the geometry of the system or extrinsically due to their location along the dusty Galactic plane. Note, however,
that the highest value of Galactic $N_{\mathrm{H}}$ is $\sim3\cdot 10^{22}$ cm$^{-2}$ so objects
with very large $N_{\mathrm{H}}$ can not be explained by interstellar absorption alone. Also keep in
mind that the absorption from \citet{Dic90} tends to be underestimated given that local
small-scale inhomogeneities and the contribution from molecular hydrogen are ignored.
The main reason that more absorbed sources are being found is that by operating above 20 keV, ISGRI
is immune to the absorption that prevented their discovery with earlier soft X-ray telescopes.
A large absorption is also a common feature of extragalactic IGRs. However, our data show that
as a group, they are not more absorbed than pre-\emph{INTEGRAL} AGN, in agreement with the conclusions of
\citet{Bec06a} and \citet{Saz07}.

Figure\,\ref{fov_disp_nh} presents an all-sky map of sources detected by ISGRI
with symbol sizes proportional to reported column densities ($N_{\mathrm{H}}$). Contours
of expected line-of-sight absorption \citep{Dic90} are provided for
levels of $10^{21}$, $5\cdot 10^{21}$ and $10^{22}$ cm$^{-2}$.
One of the benefits of such a map is that
potential clustering or asymmetries in the local distribution of matter can be studied.
The lower portion of Fig.\,\ref{fov_disp_nh} shows that the Norma Arm region hosts many of the most
heavily-absorbed Galactic sources ($N_{\mathrm{H}}\geq 10^{23}$ cm$^{-2}$) continuing a previously
noted trend \citep[e.g.,][]{Kuu05,Lut05,Wal06}. This region also happens to be the most
active formation site of young supergiant stars \citep{Bro96}. These stars are the precursors to the
absorbed HMXBs that ISGRI is discovering in the Norma Arm. The Galactic Bulge and the
Scutum/Sagittarius Arms are also represented by obscured sources but to a lesser extent than in the
Norma Arm.

For sources whose distance are known, we did not find any clear dependence of the intrinsic
$N_{\mathrm{H}}$ on the distance to the source, nor did we find a dependence of
$N_{\mathrm{H}}$ with the luminosity as derived from the soft-band fluxes (20--40 keV) listed in
\citet{Bir07}.

\subsection{Modulations}

The strong magnetic fields in some NS X-ray binaries can produce non-spherically symmetric patterns
of emission. If the magnetic and rotation axes are misaligned, this results in pulsations
in the X-ray light curve.

\begin{figure}[!t] \centering
\includegraphics[width=8cm,angle=0]{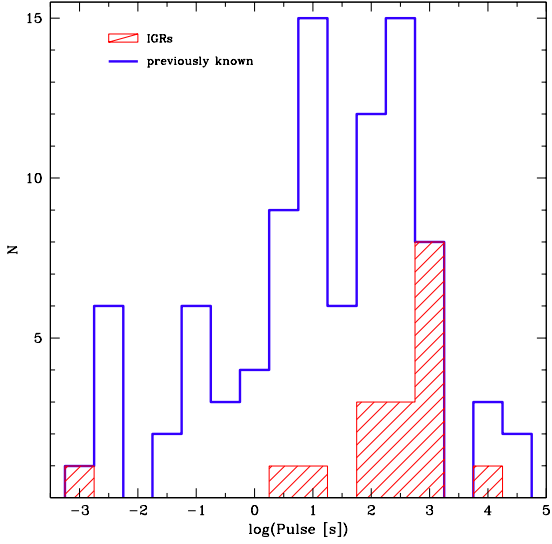}
\caption{Spin periods reported for sources detected by ISGRI that were previously known (92, clear histogram)
and for IGRs (18, shaded histogram).}
\label{hist_pulse}
\end{figure}

Most IGRs for which a pulsation has been measured have spin periods ($P_{\mathrm{s}}$) in the range of 100--1000 s, or around 10 times longer than the
average pulse period of pre-\emph{INTEGRAL} sources (Fig.\,\ref{hist_pulse}). There are notable IGRs that represent extreme cases: IGR J00291$+$5934 has a pulse
period of only 1.7 ms making it the fastest accretion-powered pulsar ever observed \citep{Gal05}, whereas IGR J16358$-$4726 has a spin period
as long as 6000 s \citep{Lut05,Pat06}. One of the reasons that IGRs have longer pulse periods than average
is because many of them are SG HMXBs which are wind-fed systems with strong magnetic fields that tend to have
the longest pulse periods \citep[e.g.][]{Cor84}. Another reason is that \emph{INTEGRAL} and \emph{XMM-Newton} feature long orbital
periods around the Earth. This means that the source can be observed for long periods of time without
interruptions, so that pulsations on the order of a few hundreds of seconds or more can be detected.
Meanwhile, the previously-known sources in Fig.\,\ref{hist_pulse} include millisecond pulsars and other LMXBs, radio pulsars, CVs, etc., which are
underrepresented among IGRs. To illustrate this, we performed a KS test which returned a very low probability (0.0007\%) of statistical compatibility
between the distributions of 18 IGRs (shaded histogram) and 92 previously-known pulsars of all types (clear histogram) as they are presented in
Fig.\,\ref{hist_pulse}. The KS-test probability improved by an order of magnitude when IGRs were compared to 49 previously-known HMXBs, and it improved by
3 orders of magnitude when IGRs were compared to 14 previously-known SG HMXBs. So \emph{INTEGRAL} is not just finding new
pulsars that are HMXBs, but these HMXBs are predominantly long-period systems with SG companions.

\begin{figure}[!t] \centering
\includegraphics[width=8cm,angle=0]{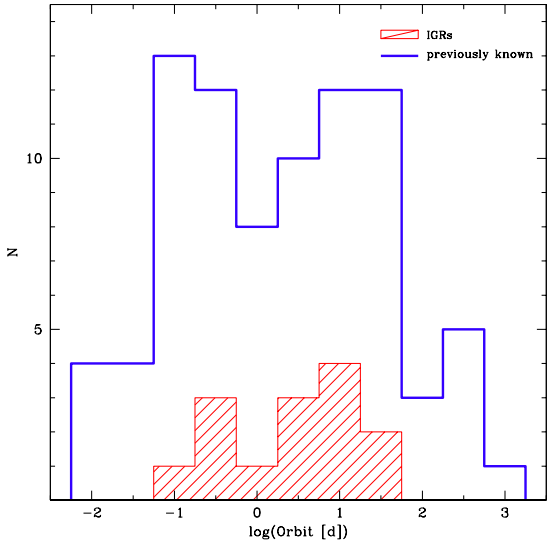}
\caption{Published orbital periods of sources detected by ISGRI. The clear histogram
represents sources that were previously known (84) while the shaded histogram represents IGRs (14).}
\label{hist_orbit}
\end{figure}

\begin{figure}[htb] \centering
\includegraphics[width=8cm,angle=0]{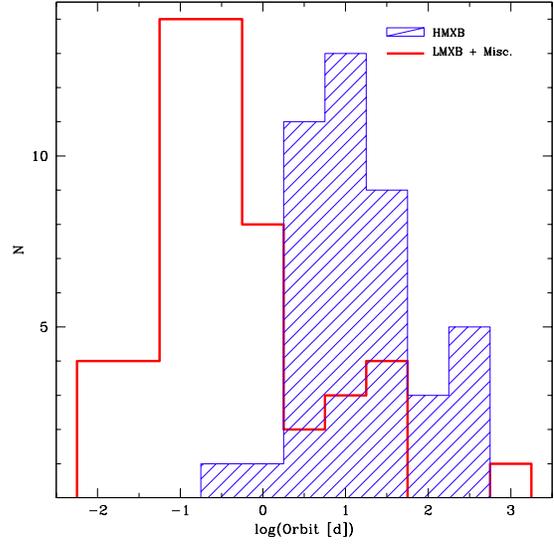}
\caption{Distribution of orbital periods of HMXBs (43, shaded histogram) compared with LMXBs
and Miscellaneous sources (54, clear histogram).}
\label{hist_orbit_hmxb}
\end{figure}

The distribution of orbital periods ($P_{\mathrm{o}}$) of IGRs exhibits a similar bimodal shape to that seen in the distribution of orbital periods known before
\emph{INTEGRAL} (Fig.\,\ref{hist_orbit}). The probability of statistical compatibility is nearly 80\%
according to a KS test. The bimodal distribution represents 2 underlying populations: LMXBs (and
miscellaneous sources) which tend to have short orbital periods, and HMXBs which tend to have longer
orbital periods (Fig.\,\ref{hist_orbit_hmxb}).

\begin{figure}[!t] \centering
\includegraphics[width=8cm,angle=0]{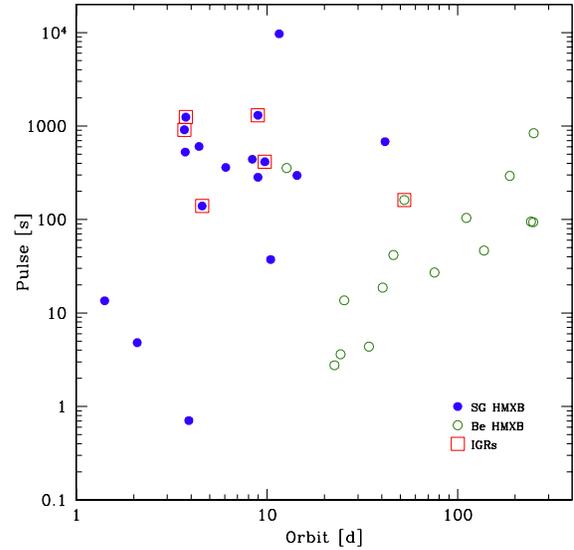}
\caption{Corbet diagram of spin vs. orbital period of HMXBs detected by ISGRI
whose companions are OB supergiants (17, filled circles) or Be stars (15, empty circles).
IGRs are boxed.}
\label{pulse_orbit}
\end{figure}

In a Corbet $P_{\mathrm{s}}$--$P_{\mathrm{o}}$ diagram \citep{Cor84}, members of each subclass of HMXBs segregate into different regions of the plot
owing to the complex feedback processes between the modulation periods and the dominant accretion mechanism. Figure\,\ref{pulse_orbit} shows
that the majority of IGRs are located among other known SG HMXBs. The figure also shows that Be HMXBs have longer orbital periods than SG HMXBs,
in general. While this fact was already known \citep[e.g.][]{Ste86}, the discrepancy remains even though \emph{INTEGRAL} has
nearly doubled the number of such systems.

\subsection{Modulations vs. Absorption}

Accretion affects the spin period of a NS. If the velocity at the corotation radius (the radius at which
the magnetic field regulates the motion of matter) exceeds the Keplerian velocity, then material will be spun away taking
angular momentum with it and the NS will slow down due to the ``propellor mechanism'' \citep{Ill75}. For corotation velocities smaller than the
Keplerian velocity, the material is able to accrete onto the NS magnetosphere which will either spin up or spin down the NS depending on
whether the angular momentum of the accreted material has the same or an opposite direction as the NS spin \citep{Wat89}. So the spin
rate of the pulsar in a HMXB is regulated by, among other things, the angular momentum of the wind of the stellar companion.

Assuming spherically-symmetric accretion from a radiation-driven wind of a SG star,
the density of the wind as a function of radius is $\rho(r) \propto r^{-2}$. On the other hand, the structure
of the winds of Be stars is believed to consist of dense slow equatorial outflows and thin fast polar winds \citep{Lam87}. The density drops
much faster with the radius ($\rho(r) \propto r^{-3}$) \citep{Wat88}. Therefore, the winds of Be stars present stronger density and velocity
gradients inside the capture radius of the NS, in both radial and azimuthal directions, which suggests that wind-fed accretion is more
efficient at delivering angular momentum to the NS in Be HMXBs than it is in SG HMXBs \citep{Wat89}.

Given the density structures described above, and assuming
a steady accretion rate of material whose angular momentum has the same direction as the spin of the NS, the spin period of the NS will
reach an equilibrium value $P_{\mathrm{eq}} \propto \rho^{-3/7}$. However, the present-day spin periods of NS in SG systems
are much longer than predicted and are actually closer to $P_{\mathrm{eq}}$ of the stellar winds while the star was still on the MS \citep{Wat89}.
The equilibrium spin period in Be systems is constantly adjusting to the changing conditions in the winds \citep{Wat89}. As with the SG systems,
pulsars in Be systems are not currently spinning at $P_{\mathrm{eq}}$ but reflect the values of an earlier evolutionary stage \citep{Kin91}.
So even though the transport of positive angular momentum through the wind is so inefficient that it can not spin up the pulsar to its
expected equilibrium spin period, this does not influence how well the pulsar can be spun down by the ``propellor mechanism'' \citep{Wat89}.

\begin{figure}[!t] \centering
\includegraphics[width=8cm,angle=0]{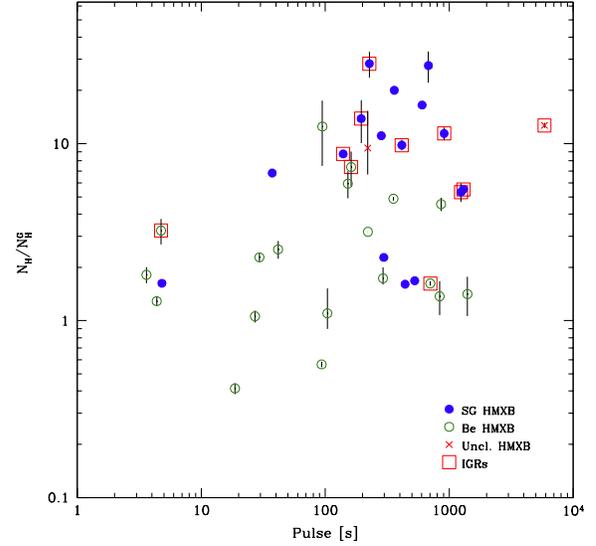}
\caption{Spin period as a function of reported $N_{\mathrm{H}}$ value (normalised by the expected Galactic value from
\citep{Dic90}) for HMXBs detected by ISGRI
whose companions are OB supergiants (16, filled circles),
Be stars (19, empty circles), or unclassified (2, crosses). IGRs are boxed and Magellanic
Cloud sources are excluded.}
\label{nh_pulse}
\end{figure}

With a few exceptions, HMXBs from the Milky Way that have been detected by ISGRI are segregated into distinct regions of a
$P_{\mathrm{s}}$--$N_{\mathrm{H}}$ diagram (Fig.\,\ref{nh_pulse}) stemming from the higher average $N_{\mathrm{H}}$ and longer
average $P_{\mathrm{s}}$ of SG HMXBs compared to Be HMXBs. The SG HMXBs set apart from the others ($P_{\mathrm{s}} < 50$ s) are
Cen X-3 which is a Roche-lobe overflow system, and OAO 1657$-$415 which might be transitioning from a wind-fed to a disk-fed system \citep{Aud06}.
The $N_{\mathrm{H}}$ values of sources in Fig.\,\ref{nh_pulse} have been normalised by the line-of-sight values ($N_{\mathrm{H}}^{\mathrm{G}}$) from \citet{Dic90}.
This normalisation does not affect our conclusions but it helps to reduce the scatter in the vertical direction,
particularly for nearby sources such as X Per.

There could be a weak positive correlation between the $N_{\mathrm{H}}$ and spin period for
HMXBs as a group. There are no highly-absorbed sources ($N_{\mathrm{H}} > 10^{23}$ cm$^{-2}$) with spin periods shorter than a few tens of seconds,
and there are no pulsars with $P_{\mathrm{s}} > 100$ s that are poorly absorbed ($N_{\mathrm{H}} < 10^{22}$ cm$^{-2}$).
A least-squares fit to the data yields $P_{\mathrm{s}} \propto N_{\mathrm{H}}^{5/7}$.
If we consider the $N_{\mathrm{H}}$ to be a reliable estimate of the density of matter around the compact object, then the slope that we find
contradicts the slope expected from the equilibrium values ($\sim -3/7$). However, as noted above,
the pulsars in Fig.\,\ref{nh_pulse} are spinning at periods that are longer than their
equilibrium values would suggest.

\begin{figure}[!t] \centering
\includegraphics[width=8cm,angle=0]{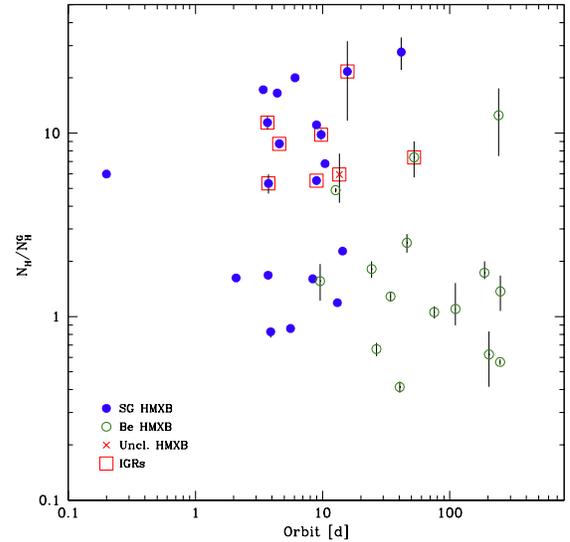}
\caption{Orbital period versus reported $N_{\mathrm{H}}$ value (normalised by the expected Galactic value from
\citep{Dic90}) for HMXBs detected by ISGRI whose companions are OB supergiants (20, filled circles),
Be stars (15, empty circles) or unclassified (1, cross). IGRs are boxed and Magellanic Cloud
sources are excluded.}
\label{nh_orbit}
\end{figure}

Since Be HMXBs tend to have longer orbital periods than SG HMXBs (see Fig.\,\ref{pulse_orbit}), a distinction is also seen among the
distributions of the $N_{\mathrm{H}}$ values and orbital periods of HMXBs with Be or SG companions (Fig.\,\ref{nh_orbit}).
There also appears to be an anti-correlation of $N_{\mathrm{H}}$ and orbital period: a least-squares fit to the data returns
$P_{\mathrm{o}} \propto N_{\mathrm{H}}^{-3/7}$.
In both types of systems, a shorter orbital period implies a compact object that is embedded deeper or spends more time in the dense regions of its stellar
companion's wind resulting in more absorption. Therefore, Be HMXBs continue the trend set by SG HMXBs into long-orbital periodicity and low-$N_{\mathrm{H}}$
regions of the plot.

Spearman rank tests to the $P_{\mathrm{s}}$--$N_{\mathrm{H}}$ and $P_{\mathrm{o}}$--$N_{\mathrm{H}}$ distributions return weak positive and negative
correlations with coefficients of 0.37 and $-$0.33, respectively, suggesting that the null hypothesis of mutual independence between $N_{\mathrm{H}}$
and $P_{\mathrm{s}}$ or $P_{\mathrm{o}}$ can be rejected. From Monte Carlo simulations, we determined that
the probability of finding a Spearman rank coefficient $\gtrsim$ 0.33 is around 5\%. Admittedly, the scatter in the data is large as can be seen in
Figs.\,\ref{nh_pulse}--\ref{nh_orbit}. Because there are large uncertainties in the $N_{\mathrm{H}}$ and practically no uncertainty in the
spin and orbital periods, the slope from a least-squares fit will tend to overestimate the real slope. Futhermore, the conclusions that we derive
for how divergent species of objects react to changes in the local absorbing matter are based on a simplification of the underlying physics.
The inclinations of the systems and their eccentricities, for example, are ignored.
Even if we can not fit a slope of $-3/7$ to the
data in Fig.\,\ref{nh_pulse}, the correlations that we find in Figs.\,\ref{nh_pulse}--\ref{nh_orbit} might
simply be due to the segregation of the 2 populations into distinct regions of the plots, rather than being due
to physical processes.

\begin{figure}[!t] \centering
\includegraphics[width=8cm,angle=0]{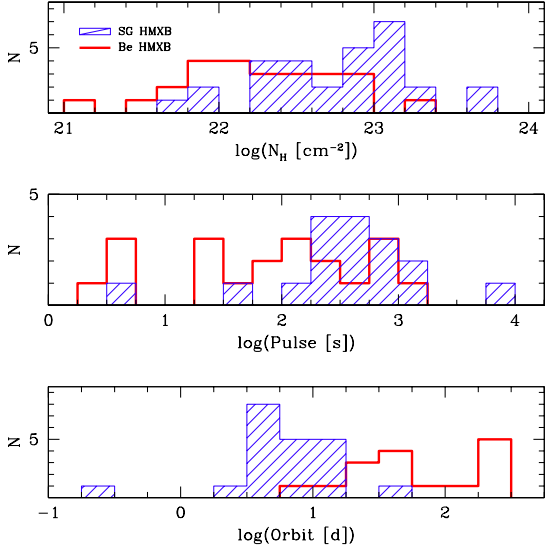}
\caption{Distribution of reported $N_{\mathrm{H}}$ (top), pulse periods (middle), and
orbital periods (bottom) among HMXBs detected by ISGRI whose companions are
OB supergiants (shaded histograms) or Be stars (clear histograms). Magellanic Cloud
sources are excluded.}
\label{hist_hmxb}
\end{figure}

Nevertheless, as more sources are added to these diagrams, the potential trends that have emerged could help constrain models describing the influence of local absorbing matter
on the modulations. Another advantage of these plots is that the probable designation of an unidentified
source is much more likely to be correct than when only a single parameter is used. This is illustrated in
Fig.\,\ref{hist_hmxb} where distributions of the 3 parameters in question ($N_{\mathrm{H}}$, $P_{\mathrm{s}}$ and
$P_{\mathrm{o}}$) are presented for SG and Be HMXBs. Other than in the orbital periods, and in the
extremities of the $N_{\mathrm{H}}$ and $P_{\mathrm{s}}$ distributions, there is little that differentiates
the 2 groups. A HMXB selected from an average bin in either $N_{\mathrm{H}}$ or $P_{\mathrm{s}}$ has a roughly equal probability of hosting a SG or Be star.
However, a random HMXB in the $N_{\mathrm{H}}$--$P_{\mathrm{s}}$ plot will tend to be located among
other members of its group.

Therefore, these diagrams could serve as new tools to help distinguish between SG and Be HMXBs when only $N_{\mathrm{H}}$ and either
the spin or orbital periods are known. For example, IGR J19140$+$0951 has an orbital period of around 13 days and $N_{\mathrm{H}} \sim 10^{23}$ cm$^{-2}$. It is positioned
among other SG HMXBs so its companion is probably an OB supergiant (boxed cross in Fig.\,\ref{nh_orbit}).
This designation has already been suggested based on other criteria such as the source's persistent emission \citep{Rod05}. Recent
IR observations of the source indicate a spectral type of B0.5-Ia which confirms the
supergiant nature of the companion (D. Hannikainen, private communication).
Similarly, AX J1749.2$-$2725 which is currently an unclassified
HMXB could have a SG companion based on its position in
Fig.\,\ref{nh_pulse} ($N_{\mathrm{H}} > 10^{23}$ cm$^{-2}$ and pulse period over 100 s).
The other unclassified HMXB in Fig.\,\ref{nh_pulse} is IGR J16358$-$4726, whose NS has an
unusually long spin period suggesting a magnetar nature for the source \citep{Pat06}. This object is
clearly in the SG HMXB camp based on its position in the plot.

\section{Summary \& Conclusions}
\label{conc}

We have compiled a catalogue of all $\sim$500 sources that were detected by ISGRI during its first 4 years of observations. This
includes published parameters such as positions, column densities, spin and orbital periods, and distances or redshifts.
The primary aims of this catalogue were to gather in a single place the most important parameters of high-energy sources
detected by ISGRI and to use this large sample to test against various theoretical predictions, to search for possible trends in the data,
and to determine where new sources fit in the parameter space established by previously-known high-energy sources.

Clustered towards the spiral arm tangents and at low Galactic latitudes, HMXBs follow the
distributions of tracers of star-forming regions. In contrast, most LMXBs are found in the
Galactic bulge or have had time to migrate to high latitudes, typical of an older stellar
population. The discrepancy is seen again in galactocentric profiles where the number of LMXBs
gradually decreases from its maximum in the central kpcs, while HMXBs avoid the central kpcs and
are overrepresented at the peaks of H\,II/CO distributions.

Over 200 new sources have been discovered by ISGRI but many of them remain unclassified.
Although some may be AGN behind the plane, unclassified sources have a spatial distribution that resembles a
Galactic population (notably LMXBs and CVs) rather than an extragalactic one. If the unclassified sources are
composed primarily of LMXBs, as their spatial distributions and the transient emission of most them seem to
suggest, then the reason they remain unclassified is because the faint optical/IR counterparts of such
sources are difficult to identify in the crowded and obscure Galactic plane.

Since it operates above 20 keV and is unhindered by absorption, ISGRI is discovering many
new HMXBs and AGN that are intrinsically absorbed ($N_{\mathrm{H}} \sim 10^{22}$--$10^{24}$ cm$^{-2}$).
On average, Galactic IGRs are more absorbed (by a factor of $\sim$4) than sources that were previously known.

Spin periods for most IGR pulsars are between a few hundred to a few thousand seconds or somewhat longer than the
average spin periods of sources known before \emph{INTEGRAL}. The distribution of orbital periods for IGRs closely resembles the
bimodal distribution set by previously-known sources. The peaks correspond to
2 underlying populations: LMXBs and miscellaneous sources such as CVs and SNRs
which tend to have short orbital periods, and HMXBs which have longer orbital periods, in general.
Almost all IGRs for which both spin and orbital periods have been measured are located in the region of wind-fed accretion in
the Corbet diagram. This is a testament to the number of new SG HMXBs that \emph{INTEGRAL} has discovered.

Thanks to the larger sample size of these new SG HMXBs, we were able to test for dependences of the
spin and orbital periods of HMXBs on the amount of absorbing matter local to the source. While scatter
is an issue, there is a clear segregation of HMXBs in both plots which could be used to help assign Be or SG
sompanions to sources that are still unclassified. There could be trends in both the
$P_{\mathrm{s}}$--$N_{\mathrm{H}}$ and $P_{\mathrm{o}}$--$N_{\mathrm{H}}$ diagrams.
The possible correlation of $P_{\mathrm{s}} \propto N_{\mathrm{H}}^{5/7}$ appears
to contradict the expected slope ($-3/7$, e.g. \citep{Cor84}) which confirms that current spin periods
are longer than the predicted equilibrium values, and that the spin-up of the pulsar via the wind is not as
effective as the spin-down via the ``propellor mechanism.'' The potential anti-correlation in the
$P_{\mathrm{o}}$--$N_{\mathrm{H}}$ plot means that the average column density varies inversely with the
distance between the objects as one would expect. However, intrinsic absorption values can change and the
potential trends we see, rather than being due to physical processes that make the parameters
inter-dependent, could simply be the result of 2 populations of sources occupying different
parameter spaces, i.e. SG HMXBs are generally more absorbed, they spin slower, and they have
shorter orbits than Be HMXBs. Nevertheless, given $N_{\mathrm{H}}$ and either
$P_{\mathrm{s}}$ or $P_{\mathrm{o}}$ for a HMXB, improves the chances of correctly predicting
the type of counterpart it has, compared with relying on only a single parameter. Of course, confirmation
of the spectral type of the donor star in a HMXB still requires an optical/IR observation.

This work takes advantage of multi-wavelength observations in order to understand the nature of IGRs, and to help clarify the
mechanisms that govern each type of source. Among the challenges facing more detailed population studies is the limited
sample size of each subclass. This can only be alleviated by using large-FOV instruments such as \emph{INTEGRAL} to search for new sources, and
by regularly observing each new source in other wavelengths so that the accumulation of evidence rules out all but a single type of object.
Many of the new sources which have been classified are absorbed
HMXBs with supergiant companions. The increasing number of these systems discovered by \emph{INTEGRAL} could alter our view of the Galactic population
of hard X-ray sources and the evolutionary scenarios of their massive stellar companions.

\begin{acknowledgements}
The authors thank the anonymous referee for their prompt review of the paper.
AB thanks S.E. Shaw, S. Paltani and M. T\"urler for their input and discussions. AB also thanks
R. Walter for useful discussions on the Galactic distribution of HMXBs. JR warmly thanks
P. Laurent, A. Goldwurm and C. Gouiffes for a careful reading of the manuscript and useful comments.
This publication uses observations obtained with the
ESA science mission \emph{INTEGRAL}. The \emph{INTEGRAL} instrument and data centres were directly funded by
ESA member states and the USA (NASA).
This research has made use of: the SIMBAD database operated at
CDS, Strasbourg, France; NASA's Astrophysics Data System Bibliographic Services;
data obtained from the High Energy Astrophysics Science Archive Research Center (HEASARC) provided
by NASA's Goddard Space Flight Center.
\end{acknowledgements}

\bibliographystyle{aa}
\bibliography{bodaghee.bib}

\clearpage

\pagestyle{empty}
\setlength{\voffset}{+0.3in}
\setlength{\textwidth}{7.2in}
\begin{tiny}
\longtabL{1}{
\label{tab_par}
\begin{landscape}
\begin{longtable}{llllllllllll}
\caption{ The parameters of sources detected by ISGRI.} \\
\hline
\noalign{\smallskip}
Name & R.A. & Dec. & Error & \emph{l} & \emph{b} & $N_{\mathrm{H}}$ & Spin & Orbit & Distance & Type & Ref.  \\
 & \multicolumn{2}{c}{(J2000)}  & (min) & \multicolumn{2}{c}{(deg)} & (10$^{22}$ cm$^{-2}$) & (s) & (d) & (kpc or [z]) &  &  \\
\noalign{\smallskip}
\noalign{\smallskip}
\hline
\hline
\noalign{\smallskip}
\noalign{\smallskip}
\endfirsthead
\caption{ continued.} \\
\hline
\noalign{\smallskip}
Name & R.A. & Dec. & Error & \emph{l} & \emph{b} & $N_{\mathrm{H}}$ & Spin & Orbit & Distance & Type & Ref.  \\
 & \multicolumn{2}{c}{(J2000)}  & \multicolumn{3}{c}{(deg)} & (10$^{22}$ cm$^{-2}$) & (s) & (d) & (kpc or [z]) &  &  \\
\noalign{\smallskip}
\noalign{\smallskip}
\hline
\hline
\noalign{\smallskip}
\noalign{\smallskip}
\endhead
\endfoot

IGR J00040$+$7020 & 00 04 01 & $+$70 20 10 & 3.8 & 118.930 & 7.834 & -- & -- & -- & -- & Unclassified (AGN?) & 1 \\ 
\noalign{\smallskip}
IGR J00234$+$6141 & 00 22 54 & $+$61 42 22 & 4.8 & 119.557 & $-$0.978 & -- & 570 & -- & 0.3 & CV (IP) & 1,2,3 \\ 
\noalign{\smallskip}
IGR J00245$+$6251 & 00 24 28 & $+$62 50 35 & 2.3 & 119.860 & 0.132 & -- & -- & -- & -- & GRB & 1 \\ 
\noalign{\smallskip}
4U 0022$+$63 & 00 25 17 & $+$64 09 32 & 3.6 & 120.086 & 1.432 & -- & -- & -- & 2.83(79) & SNR & 1,4 \\ 
\noalign{\smallskip}
IGR J00254$+$6822 & 00 25 26 & $+$68 22 50 & 2.3 & 120.529 & 5.630 & -- & -- & -- & -- & Unclassified (Sey-2?) & 5 \\ 
\noalign{\smallskip}
V709 Cas & 00 28 48.87 & $+$59 17 21.8 & 0.017 & 120.042 & $-$3.455 & 0.8$\pm$0.4 & 312.746(3) & 0.2225(2) & 0.230(20) & CV (DQ Her) & 6,7,8,9 \\ 
\noalign{\smallskip}
IGR J00291$+$5934 & 00 29 03.08 & $+$59 34 19.2 & 0.01 & 120.096 & $-$3.176 & 0.43$_{-0.05}^{+0.07}$ & 0.0016698(1) & 0.1023622(4) & 5.1$_{-0.4}^{+0.3}$ & LMXB (P, T) & 10,11,12,13,14 \\ 
\noalign{\smallskip}
IGR J00335$+$6126 & 00 33 35 & $+$61 26 52 & 3.3 & 120.800 & $-$1.350 & -- & -- & -- & -- & Unclassified & 5 \\
\hline
\end{longtable}
\hspace{-6mm} \textbf{Name}: One commonly used name for the source \\
\textbf{R.A.}: Right Ascension (J2000) in \emph{hh mm ss.ss} \\
\textbf{Dec.}: Declination in \emph{dd mm ss.ss} \\
\textbf{Error}: Error radius in minutes \\
\textbf{\emph{l}}: Galactic longitude in degrees \\
\textbf{\emph{b}}: Galactic latitude in degrees \\
\textbf{$N_{\mathrm{H}}$}: Column density in 10$^{22}$ cm$^{-2}$ \\
\textbf{Spin}: Spin period in seconds \\
\textbf{Orbit}: Orbital period in days \\
\textbf{Distance}: Distance in kpc for galactic sources (including LMC/SMC), or redshift (in brackets) for extragalactic sources \\
\textbf{Type}: A (atoll), AGN (active galactic nucleus), AXP (anomalous X-ray pulsar), B (burster),
Be (Be star), BHC (black hole candidate), CV (cataclysmic variable), E (eclipsing), D (dipping),
DN (dwarf nova), F (flaring), GRS (gamma-ray source), HMXB (high-mass X-ray binary),
IP (intermediate polar), LMXB (low-mass X-ray binary), Mol. Cloud (molecular cloud), muQSO (micro-quasar), N (nova), P (pulsar),
PWN (pulsar wind nebula), QPO (quasi-periodic oscillations), QSO (quasar),
RP (radio pulsar), Sey (Seyfert galaxy),
SFXT (supergiant fast X-ray transient), SG (OB supergiant), SGR (soft gamma repeater),
SNR (supernova remnant), Symb (symbiotic star), T (transient), Z (Z-track) \\
\textbf{References}: References for the listed parameters \\
\end{landscape}

}

\end{tiny}

\setlength{\voffset}{0in}
\setlength{\textwidth}{7in}

\onecolumn
\noindent [1] Bird A.J., Malizia A., Bazzano A., et al., 2007, \apj, in press, astro-ph/0611493

\noindent [2] Bikmaev I.F., Revnivtsev M.G., Burenin R.A., \& Sunyaev R.A., 2006, AstL, 32, 588

\noindent [3] Masetti N., Bassani L., Bazzano A., et al., 2006, \aap,  455, 11

\noindent [4] Ruiz-Lapuente P., 2004, \apj, 612, 357

\noindent [5] Kuiper L., Hartog P.R.D., \& Hermsen W., 2006, ATel, 939, 1

\noindent [6] Downes R., Webbink R.F., \& Shara M.M., 1997, \pasp, 109, 345

\noindent [7] de Martino D., Matt G., Mukai K., et al., 2001, \aap, 377, 499

\noindent [8] Bonnet-Bidaud J.M., Mouchet M., de Martino D., et al., 2001, \aap, 374, 1003

\noindent [9] Barlow E.J., Knigge C., Bird A.J., et al., 2006, \mnras, 372, 224

\noindent [10] Nowak M.A., Paizis A., Wilms J., et al., 2004, ATel, 369, 1

\noindent [11] Paizis A., Nowak M.A., Wilms J., et al., 2005, \aap, 444, 357

\noindent [12] Burderi L., di Salvo T., Riggio A., et al., 2006, ChJAS, 6, 192

\noindent [13] Galloway D.K., Markwardt C.B., Morgan E.H., et al., 2005, \apjl, 622, L45

\noindent [14] Falanga M., Kuiper L., Poutanen J., et al., 2005, \aap, 444, 15

\noindent [15] Laurent-Muehleisen S.A., Kollgaard R.I., Ryan P.J., et al., 1997, \aaps, 122, 235

\noindent [16] Bassani L., Molina M., Malizia A., et al., 2006, \apjl, 636, L65

\noindent [17] Perryman M.A.C., Lindegren L., Kovalevsky J., et al., 1997, \aap, 323, L49

\noindent [18] den Hartog P.R., Hermsen W., Kuiper L., et al., 2006, \aap, 451, 587

\noindent [19] Wen L., Levine A.M., Corbet R.H.D., \& Bradt H.V., 2006, \apjs, 163, 372

\noindent [20] Negueruela I. \& Reig P., 2004, ATel, 285, 1

\noindent [21] Snellen I.A.G., McMahon R.G., Hook I.M., \& Browne I.W.A., 2002, \mnras, 329, 700

\noindent [22] Sazonov S., Revnivtsev M., Krivonos R., et al., 2007, \aap, 462, 57

\noindent [23] Yu P.-C. \& Hwang C.-Y., 2005, \apj, 631, 720

\noindent [24] Liu Q.Z., van Paradijs J., \& van den Heuvel E.P.J., 2000, \aaps, 147, 25

\noindent [25] Buckley D.A.H., Coe M.J., Stevens J.B., et al., 2001, \mnras, 320, 281

\noindent [26] Corbet R., Marshall F.E., Lochner J.C., et al., 1998, \iaucirc, 6803, 1

\noindent [27] Galache J.L., Corbet R.H.D., Coe M.J., et al., 2005, ATel, 674, 1

\noindent [28] Smith M.A., Cohen D.H., Gu M.F., et al., 2004, \apj, 600, 972

\noindent [29] Harmanec P., Habuda P., \v Stefl S., et al., 2000, \aap, 364, L85

\noindent [30] Grimm H.-J., Gilfanov M., \& Sunyaev R., 2002, \aap, 391, 923

\noindent [31] Massey P., 2002, \apjs, 141, 81

\noindent [32] Naik S. \& Paul B., 2004, \aap, 418, 655

\noindent [33] Corbet R.H.D., Finley J.P., \& Peele A.G., 1999, \apj, 511, 876

\noindent [34] Crampton D., Hutchings J.B., \& Cowley A.P., 1985, \apj, 299, 839

\noindent [35] Reig P., Chakrabarty D., Coe M.J., et al., 1996, \aap, 311, 879

\noindent [36] Campana S., Gastaldello F., Stella L., et al., 2001, \apj, 561, 924

\noindent [37] Tamura K., Tsunemi H., Kitamoto S., et al., 1992, \apj, 389, 676

\noindent [38] Negueruela I. \& Okazaki A.T., 2001, \aap, 369, 108

\noindent [39] da Costa L.N., Willmer C.N.A., Pellegrini P.S., et al., 1998, \aj, 116, 1

\noindent [40] Risaliti G., 2002, \aap, 386, 379

\noindent [41] Polletta M., Bassani L., Malaguti G., et al., 1996, \apjs, 106, 399

\noindent [42] Reig P., Negueruela I., Papamastorakis G., et al., 2005, \aap, 440, 637

\noindent [43] Loveday J., 1996, \mnras, 278, 1025

\noindent [44] Hewitt A. \& Burbidge G., 1991, \apjs, 75, 297

\noindent [45] Hulleman F., van Kerkwijk M.H., \& Kulkarni S.R., 2004, \aap, 416, 1037

\noindent [46] G\"ohler E., Wilms J., \& Staubert R., 2005, \aap, 433, 1079

\noindent [47] Pigulski A., Kopacki G., \& Ko{\l}aczkowski Z., 2001, \aap, 376, 144

\noindent [48] Mereghetti S., Tiengo A., Israel G.L., \& Stella L., 2000, \aap, 354, 567

\noindent [49] Burenin R., Mescheryakov A., Sazonov S., et al., 2006, ATel, 883, 1

\noindent [50] Den Hartog P.R., Kuiper L., Hermsen W., et al., 2005, ATel, 394, 1

\noindent [51] Kennea J.A., Racusin J.L., Burrows D.N., et al., 2005, ATel, 673, 1

\noindent [52] Argyle R.W. \& Eldridge P., 1990, \mnras, 243, 504

\noindent [53] Kuiper L., Hermsen W., in 't Zand J., \& den Hartog P.R., 2005, ATel, 665, 1

\noindent [54] Cotton W.D., Condon J.J., \& Arbizzani E., 1999, \apjs, 125, 409

\noindent [55] Pfefferkorn F., Boller T., \& Rafanelli P., 2001, \aap, 368, 797

\noindent [56] Paturel G. \& Petit C., 2002, LEDA (2002), 0

\noindent [57] Clements E.D., 1981, \mnras, 197, 829

\noindent [58] Krongold Y., Nicastro F., Elvis M., et al., 2005, \apj, 620, 165

\noindent [59] Arribas S., Mediavilla E., del Burgo C., \& Garc\'ia-Lorenzo B., 1999, \apj, 511, 680

\noindent [60] Leahy D.A., Harrison F.A., \& Yoshida A., 1997, \apj, 475, 823

\noindent [61] Taylor A.R. \& Gregory P.C., 1982, \apj, 255, 210

\noindent [62] Ma C., Arias E.F., Eubanks T.M., et al., 1998, \aj, 116, 516

\noindent [63] Weaver K.A., Wilson A.S., Henkel C., \& Braatz J.A., 1999, \apj, 520, 130

\noindent [64] Fischer J.-U., Hasinger G., Schwope A.D., et al., 1998, AN, 319, 347

\noindent [65] Cappi M., Panessa F., Bassani L., et al., 2006, \aap, 446, 459

\noindent [66] Titov O.A., 2004, ARep, 48, 941

\noindent [67] Karachentsev I.D., Karachentseva V.E., Kudrya Y.N., et al., 1999, Bull. Special Astrophys.Obs., 47, 5

\noindent [68] Keel W.C., 1996, \apjs, 106, 27

\noindent [69] Ishwara-Chandra C.H. \& Saikia D.J., 1999, \mnras, 309, 100

\noindent [70] Beasley A.J., Gordon D., Peck A.B., et al., 2002, \apjs, 141, 13

\noindent [71] Enya K., Yoshii Y., Kobayashi Y., et al., 2002, \apjs, 141, 23

\noindent [72] Vrielmann S., Ness J.-U., \& Schmitt J.H.M.M., 2005, \aap, 439, 287

\noindent [73] Burenin R., Mescheryakov A., Revnivtsev M., et al., 2006, ATel, 880, 1

\noindent [74] Lawrence A., Rowan-Robinson M., Ellis R.S., et al., 1999, \mnras, 308, 897

\noindent [75] Unger S.J., Norton A.J., Coe M.J., \& Lehto H.J., 1992, \mnras, 256, 725

\noindent [76] Kreykenbohm I., Mowlavi N., Produit N., et al., 2005, \aap, 433, L45

\noindent [77] Stella L., White N.E., Davelaar J., et al., 1985, \apjl, 288, L45

\noindent [78] Negueruela I., Roche P., Fabregat J., \& Coe M.J., 1999, \mnras, 307, 695

\noindent [79] di Salvo T., Burderi L., Robba N.R., \& Guainazzi M., 1998, \apj, 509, 897

\noindent [80] Delgado-Mart\'i, H., Levine A.M., Pfahl E., \& Rappaport S.A., 2001, \apj, 546, 455

\noindent [81] Lyubimkov L.S., Rostopchin S.I., Roche P., \& Tarasov A.E., 1997, \mnras, 286, 549

\noindent [82] Fomalont E.B., Frey S., Paragi Z., et al., 2000, \apjs, 131, 95

\noindent [83] Motch C., Guillout P., Haberl F., et al., 1998, \aaps, 132, 341

\noindent [84] Lauberts A., 1982, ESO/Uppsala survey of the ESO(B) atlas
(Garching: European Southern Observatory (ESO), 1982)

\noindent [85] Sazonov S., Churazov E., Revnivtsev M., Vikhlinin A., \& Sunyaev R., 2005, \aap, 444, L37

\noindent [86] Masetti N., Morelli L., Palazzi E., et al., 2006, \aap, 459, 21

\noindent [87] Remillard R.A., Bradt H.V.D., Brissenden R.J.V., et al., 1993, \aj, 105, 2079

\noindent [88] Takata T., Yamada T., Saito M., Chamaraux P., \& Kazes I., 1994, \aaps, 104, 529

\noindent [89] Kulkarni S.R., Kaplan D.L., Marshall H.L., et al., 2003, \apj, 585, 948

\noindent [90] Haberl F., Dennerl K., \& Pietsch W., 2003, \aap, 406, 471

\noindent [91] Dennerl K., Haberl F., \& Pietsch W., 1995, \iaucirc, 6184, 2

\noindent [92] G\"otz D., Mereghetti S., Merlini D., et al., 2006, \aap, 448, 873

\noindent [93] Fuhrmeister B. \& Schmitt J.H.M.M., 2003, \aap, 403, 247

\noindent [94] Naik S. \& Paul B., 2002, JApA, 23, 27

\noindent [95] Levine A.M., Rappaport S.A., \& Zojcheski G., 2000, \apj, 541, 194

\noindent [96] Han J.L. \& Tian W.W., 1999, \aaps, 136, 571

\noindent [97] Kirsch M.G.F., Sch\"onherr G., Kendziorra E., et al., 2006, \aap, 453, 173

\noindent [98] Lyne A.G., Pritchard R.S., \& Graham-Smith F., 1993, \mnras, 265, 1003

\noindent [99] Trimble V., 1973, \pasp, 85, 579

\noindent [100] Mukherjee U. \& Paul B., 2005, \aap, 431, 667

\noindent [101] Steele I.A., Negueruela I., Coe M.J., \& Roche P., 1998, \mnras, 297, L5

\noindent [102] Cui W., Feng Y.X., Zhang S.N., et al., 2002, \apj, 576, 357

\noindent [103] Levine A.M. \& Corbet R., 2006, ATel, 940, 1

\noindent [104] Haardt F., Galli M.R., Treves A., et al., 2001, \apjs, 133, 187

\noindent [105] Hirayama M., Nagase F., Endo T., et al., 2002, \mnras, 333, 603

\noindent [106] Livingstone M.A., Kaspi V.M., \& Gavriil F.P., 2005, \apj, 633, 1095

\noindent [107] Ramsay G. \& Cropper M., 2002, \mnras, 334, 805

\noindent [108] Pavlenko E.P., 2006, Astrophysics, 49, 105

\noindent [109] Matt G., Bianchi S., de Rosa A., et al., 2006, \aap, 445, 451

\noindent [110] Helou G. \& Walker D.W., eds. 1988, IRAS catalogs and atlases. Volume 7: The small scale structure catalog

\noindent [111] Kennea J.A., Markwardt C.B., Tueller J., et al., 2005, ATel, 677, 1

\noindent [112] Morelli L., Masetti N., Bassani L., et al., 2006, ATel, 785, 1

\noindent [113] Chenevez J., Budtz-Jorgensen C., Lund N., et al., 2004, ATel, 223, 1

\noindent [114] Mauch T., Murphy T., Buttery H.J., et al., 2003, \mnras, 342, 1117

\noindent [115] Brinkmann W. \& Siebert J., 1994, \aap, 285, 812

\noindent [116] Cappi M., Bassani L., Comastri A., et al., 1999, \aap, 344, 857

\noindent [117] Liu Q.Z., van Paradijs J., \& van den Heuvel E.P.J., 2001, \aap, 368, 1021

\noindent [118] Piraino S., Santangelo A., Ford E.C., \& Kaaret P., 1999, \aap, 349, L77

\noindent [119] Ford E., Kaaret P., Tavani M., et al., 1997, \apjl, 475, L123

\noindent [120] Paerels F., Brinkman A.C., van der Meer R.L.J., et al., 2001, \apj, 546, 338

\noindent [121] Araujo-Betancor S., G\"ansicke B.T., Hagen H.-J., et al., 2003, \aap, 406, 213

\noindent [122] Barkhouse W.A. \& Hall P.B., 2001, \aj, 121, 2843

\noindent [123] Malizia A., Bassani L., Capalbi M., et al., 2003, \aap, 406, 105

\noindent [124] Sidoli L., Parmar A.N., \& Oosterbroek T., 2005, \aap, 429, 291

\noindent [125] Jonker P.G. \& Nelemans G., 2004, \mnras, 354, 355

\noindent [126] Bird A.J., Barlow E.J., Bassani L., et al., 2006, \apj, 636, 765

\noindent [127] Landi R., Malizia A., Bassani L., et al., 2006, astro-ph/0610358

\noindent [128] Revnivtsev M.G., Sazonov S.Y., Molkov S.V., et al., 2006, AstL, 32, 145

\noindent [129] Porquet D., Reeves J.N., O'Brien P., \& Brinkmann W., 2004, \aap, 422, 85

\noindent [130] Pavlov G.G., Zavlin V.E., Sanwal D., et al., 2001, \apjl, 552, L129

\noindent [131] Caraveo P.A., De Luca A., Mignani R.P., \& Bignami G.F., 2001, \apj, 561, 930

\noindent [132] Belloni T., Hasinger G., Pietsch W., et al., 1993, \aap, 271, 487

\noindent [133] Kennea J.A. \& Campana S., 2006, ATel, 818, 1

\noindent [134] Grandi P., Malaguti G., \& Fiocchi M., 2006, \apj, 642, 113

\noindent [135] Orlandini M., dal Fiume D., Frontera F., et al., 1998, \aap, 332, 121

\noindent [136] Kreykenbohm I., Coburn W., Wilms J., et al., 2002, \aap, 395, 129

\noindent [137] Sadakane K., Hirata R., Jugaku J., et al., 1985, \apj, 288, 284

\noindent [138] Kirhakos S., Strauss M.A., Yahil A., et al., 1991, \aj, 102, 1933

\noindent [139] Tueller J., Barthelmy S., Burrows D., et al., 2005, ATel, 669, 1

\noindent [140] Juett A.M. \& Chakrabarty D., 2003, \apj, 599, 498

\noindent [141] Voges W., Aschenbach B., Boller T., et al., 1999, \aap, 349, 389

\noindent [142] Pineda F. \& Schnopper H.W., 1978, \iaucirc, 3190, 3

\noindent [143] Balestra I., Bianchi S., \& Matt G., 2004, \aap, 415, 437

\noindent [144] Paizis A., Gotz D., Sidoli L., et al., 2006, ATel, 865, 1

\noindent [145] Shrader C.R., Sutaria F.K., Singh K.P., \& Macomb D.J., 1999, \apj, 512, 920

\noindent [146] Ajello M., Greiner J., K\"upc\"u Yoldas A., et al., 2006, ATel, 864, 1

\noindent [147] Zacharias N. Urban S.E. Zacharias M.I. et al., 2003, VizieR Online Data Catalog, 1289

\noindent [148] Ho L.C., Filippenko A.V., \& Sargent W.L., 1995, \apjs, 98, 477

\noindent [149] Veron-Cetty M.-P. \& Veron P., 1996, A Catalogue of quasars and active nuclei
(ESO Scientific Report, Garching: European Southern Observatory (ESO), 1996, 7th ed.)

\noindent [150] Vignali C. \& Comastri A., 2002, \aap, 381, 834

\noindent [151] Reig P. \& Roche P., 1999, \mnras, 306, 100

\noindent [152] Masetti N., Pretorius M.L., Palazzi E., et al., 2006, \aap, 449, 1139

\noindent [153] Burderi L., Di Salvo T., Robba N.R., et al., 2000, \apj, 530, 429

\noindent [154] Nagase F., Corbet R.H.D., Day C.S.R., et al., 1992, \apj, 396, 147

\noindent [155] Hutchings J.B., Cowley A.P., Crampton D., et al., 1979, \apj, 229, 1079

\noindent [156] Steeghs D., Torres M.A.P., \& Jonker P.G., 2006, ATel, 768, 1

\noindent [157] Smith D.M., Bezayiff N., \& Negueruela I., 2006, ATel, 773, 1

\noindent [158] Sidoli L., Paizis A., \& Mereghetti S., 2006, \aap, 450, L9

\noindent [159] Krivonos R., Molkov S., Revnivtsev M., et al., 2005, ATel, 545, 1

\noindent [160] De Rosa A., Piro L., Fiore F., et al., 2002, \aap, 387, 838

\noindent [161] in 't Zand J. \& Heise J., 2004, ATel, 362, 1

\noindent [162] Corbet R.H.D. \& Remillard R., 2005, ATel, 377, 1

\noindent [163] Ray P.S. \& Chakrabarty D., 2002, \apj, 581, 1293

\noindent [164] Cook M.C. \& Warwick R.S., 1987, \mnras, 227, 661

\noindent [165] Stevens J.B., Reig P., Coe M.J., et al., 1997, \mnras, 288, 988

\noindent [166] Beckmann V., Gehrels N., Shrader C.R., \& Soldi S., 2006, \apj, 638, 642

\noindent [167] Becker R.H., White R.L., \& Helfand D.J., 1995, \apj, 450, 559

\noindent [168] Risaliti G., Gilli R., Maiolino R., \& Salvati M., 2000, \aap, 357, 13

\noindent [169] Li J. \& Jin W., 1996, \aaps, 120, 201

\noindent [170] La Barbera A., Segreto A., Santangelo A., et al., 2005, \aap, 438, 617

\noindent [171] Leahy D.A., 2002, \aap, 391, 219

\noindent [172] Tueller J., Gehrels N., Mushotzky R.F., et al., 2005, ATel, 591, 1

\noindent [173] Clements E.D., 1983, \mnras, 204, 811

\noindent [174] Guainazzi M., Perola G.C., Matt G., et al., 1999, \aap, 346, 407

\noindent [175] Fairall A.P., Woudt P.A., \& Kraan-Korteweg R.C., 1998, \aaps, 127, 463

\noindent [176] Torrej\'on J.M. \& Orr A., 2001, \aap, 377, 148

\noindent [177] Perryman M.A.C. \& ESA., 1997, The HIPPARCOS and TYCHO catalogues.
  Astrometric and photometric star catalogues derived from the ESA HIPPARCOS
  Space Astrometry Mission (Publisher: Noordwijk, Netherlands: ESA Publications
  Division, 1997, Series: ESA SP Series vol no: 1200, ISBN: 9290923997 (set))

\noindent [178] Jackson C.A., Wall J.V., Shaver P.A., et al., 2002, \aap, 386, 97

\noindent [179] Bassa C.G., Jonker P.G., in 't Zand J.J.M., \& Verbunt F., 2006, \aap, 446, L17

\noindent [180] Boller T., Haberl F., Voges W., Piro L., \& Heise J., 1997, \iaucirc, 6546, 1

\noindent [181] Jerjen H. \& Dressler A., 1997, \aaps, 124, 1

\noindent [182] Boirin L. \& Parmar A.N., 2003, \aap, 407, 1079

\noindent [183] in 't Zand J.J.M., Kuulkers E., Verbunt F., et al., 2003, \aap, 411, L487

\noindent [184] Mahdavi A. \& Geller M.J., 2001, \apjl, 554, L129

\noindent [185] Arnaud M., Aghanim N., Gastaud R., et al., 2001, \aap, 365, L67

\noindent [186] Chernyakova M., Lutovinov A., Rodriguez J., \& Revnivtsev M., 2005, \mnras, 364, 455

\noindent [187] Bicay M.D., Stepanian J.A., Chavushyan V.H., et al., 2000, \aaps, 147, 169

\noindent [188] Done C., Madejski G.M., \.Zycki P.T., \& Greenhill L.J., 2003, \apj, 588, 763

\noindent [189] Odewahn S.C. \& Aldering G., 1996, Private Communication, 1

\noindent [190] Palumbo G.G.C., et al., 1988, in Accurate positions of Zwicky galaxies (1988), 0--+

\noindent [191] Nilson P., 1973, Nova Acta Regiae Soc.Sci.Upsaliensis Ser.V, 0

\noindent [192] Evans D.A., Kraft R.P., Worrall D.M., et al., 2004, \apj, 612, 786

\noindent [193] Church M.J., Reed D., Dotani T., Ba{\l}uci\'nska-Church M., \& Smale A.P., 2005, \mnras, 359, 1336

\noindent [194] Parmar A.N., Gottwald M., van der Klis M., \& van Paradijs J., 1989, \apj, 338, 1024

\noindent [195] Kaldare R., Colless M., Raychaudhury S., \& Peterson B.A., 2003, \mnras, 339, 652

\noindent [196] Demircan O., Eker Z., Karata{\c s} Y., \& Bilir S., 2006, \mnras, 366, 1511

\noindent [197] Willmer C.N.A., Maia M.A.G., Mendes S.O., et al., 1999, \aj, 118, 1131

\noindent [198] Papadakis I.E., Kazanas D., \& Akylas A., 2005, \apj, 631, 727

\noindent [199] Piconcelli E., S\'anchez-Portal M., Guainazzi M., et al., 2006, \aap, 453, 839

\noindent [200] Imamura J.N., Steiman-Cameron T.Y., \& Wolff M.T., 2000, \pasp, 112, 18

\noindent [201] Wright A.E., Griffith M.R., Hunt A.J., et al., 1996, \apjs, 103, 145

\noindent [202] Matt G., Guainazzi M., Maiolino R., et al., 1999, \aap, 341, L39

\noindent [203] Bianchi S., Balestra I., Matt G., et al., 2003, \aap, 402, 141

\noindent [204] Nicastro F., Piro L., De Rosa A., et al., 2000, \apj, 536, 718

\noindent [205] Crenshaw D.M. \& Kraemer S.B., 1999, \apj, 521, 572

\noindent [206] West R.M., Surdej J., Schuster H.-E., et al., 1981, \aaps, 46, 57

\noindent [207] DeLaney T., Gaensler B.M., Arons J., \& Pivovaroff M.J., 2006, \apj, 640, 929

\noindent [208] Livingstone M.A., Kaspi V.M., Gavriil F.P., \& Manchester R.N., 2005, \apj, 619, 1046

\noindent [209] Gaensler B.M., Brazier K.T.S., Manchester R.N., et al., 1999, \mnras, 305, 724

\noindent [210] Iaria R., Di Salvo T., Burderi L., \& Robba N.R., 2001, \apj, 561, 321

\noindent [211] Robba N.R., Burderi L., Di Salvo T., et al., 2001, \apj, 562, 950

\noindent [212] Baykal A., Inam S.\c C., \& Beklen E., 2006, \aap, 453, 1037

\noindent [213] Reynolds A.P., Bell S.A., \& Hilditch R.W., 1992, \mnras, 256, 631

\noindent [214] in 't Zand J.J.M., Corbet R.H.D., \& Marshall F.E., 2001, \apjl, 553, L165

\noindent [215] Farinelli R., Frontera F., Masetti N., et al., 2003, \aap, 402, 1021

\noindent [216] Wang Z. \& Chakrabarty D., 2004, \apjl, 616, L139

\noindent [217] Stephen J.B., Bassani L., Molina M., et al., 2005, \aap, 432, L49

\noindent [218] Haberl F., Motch C., \& Zickgraf F.-J., 2002, \aap, 387, 201

\noindent [219] de Martino D., Bonnet-Bidaud J.-M., Mouchet M., et al., 2006, \aap, 449, 1151

\noindent [220] Klemola A.R., Jones B.F., \& Hanson R.B., 1987, \aj, 94, 501

\noindent [221] Kazarovets E.V., Samus N.N., \& Durlevich O.V., 2001, IBVS, 5135, 1

\noindent [222] Corbel S., Tomsick J.A., \& Kaaret P., 2006, \apj, 636, 971

\noindent [223] Orosz J.A., Groot P.J., van der Klis M., et al., 2002, \apj, 568, 845

\noindent [224] Smale A.P., 1991, \pasp, 103, 636

\noindent [225] Woudt P.A. \& Kraan-Korteweg R.C., 2001, \aap, 380, 441

\noindent [226] Greco M.V., Miller J.M., \& Steeghs D., 2006, ATel, 858, 1

\noindent [227] Wachter S., Hoard D.W., Bailyn C.D., et al., 2002, \apj, 568, 901

\noindent [228] Tomsick J.A., Chaty S., Rodriguez J., et al., 2006, \apj, 647, 1309

\noindent [229] Kaspi V.M., Crawford F., Manchester R.N., et al., 1998, \apjl, 503, L161

\noindent [230] Gotthelf E.V., Petre R., \& Hwang U., 1997, \apjl, 487, L175

\noindent [231] Torii K., Gotthelf E.V., Vasisht G., et al., 2000, \apjl, 534, L71

\noindent [232] Caswell J.L., Murray J.D., Roger R.S., et al., 1975, \aap, 45, 239

\noindent [233] Garmire G.P., Pavlov G.G., Garmire A.B., \& Zavlin V.E., 2000, \iaucirc, 7350, 2

\noindent [234] Reynoso E.M., Green A.J., Johnston S., et al., 2004, PASA, 21, 82

\noindent [235] McNamara B.J., Harrison T.E., Zavala R.T., et al., 2003, \aj, 125, 1437

\noindent [236] Bradshaw C.F., Geldzahler B.J., \& Fomalont E.B., 2003, \apj, 592, 486

\noindent [237] Vanderlinde K.W., Levine A.M., \& Rappaport S.A., 2003, \pasp, 115, 739

\noindent [238] Campana S., Belloni T., Homan J., et al., 2006, ATel, 688, 1

\noindent [239] Markwardt C.B. \& Swank J.H., 2005, ATel, 679, 1

\noindent [240] Wachter S., Wellhouse J.W., Patel S.K., et al., 2005, \apj, 621, 393

\noindent [241] Parmar A.N., Oosterbroek T., Boirin L., \& Lumb D., 2002, \aap, 386, 910

\noindent [242] Christian D.J. \& Swank J.H., 1997, \apjs, 109, 177

\noindent [243] Kennea J.A., Burrows D.N., Nousek J.A., et al., 2005, ATel, 459, 1

\noindent [244] Beckmann V., Kennea J.A., Markwardt C., et al., 2005, \apj, 631, 506

\noindent [245] Rodriguez J. \& Goldwurm A., 2003, ATel, 201, 1

\noindent [246] Schartel N., Ehle M., Breitfellner M., et al., 2003, \iaucirc, 8072, 3

\noindent [247] Walter R., Zurita Heras J.A., Bassani L., et al., 2006, \aap, 453, 133

\noindent [248] Filliatre P. \& Chaty S., 2004, \apj, 616, 469

\noindent [249] Rodriguez J., Bodaghee A., Kaaret P., et al., 2006, \mnras, 366, 274

\noindent [250] Corbet R., Barbier L., Barthelmy S., et al., 2005, ATel, 649, 1

\noindent [251] Orlandini M., Fiume D.D., Frontera F., et al., 1998, \apjl, 500, L163

\noindent [252] Owens A., Oosterbroek T., \& Parmar A.N., 1997, \aap, 324, L9

\noindent [253] Middleditch J., Mason K.O., Nelson J.E., \& White N.E., 1981, \apj, 244, 1001

\noindent [254] Tomsick J.A., Corbel S., Goldwurm A., \& Kaaret P., 2005, \apj, 630, 413

\noindent [255] Kouveliotou C., Patel S., Tennant A., et al., 2003, \iaucirc, 8109, 2

\noindent [256] Patel S.K., Zurita Heras J.A., Del Santo M., et al., 2006, \apj, in press, astro-ph/0610768

\noindent [257] Ebeling H., Mullis C.R., \& Tully R.B., 2002, \apj, 580, 774

\noindent [258] McHardy I.M., Lawrence A., Pye J.P., \& Pounds K.A., 1981, \mnras, 197, 893

\noindent [259] Bodaghee A., Walter R., Zurita Heras J.A., et al., 2006, \aap, 447, 1027

\noindent [260] Thompson T.W.J., Tomsick J.A., Rothschild R.E., et al., 2006, \apj, 649, 373

\noindent [261] Vrtilek S.D., McClintock J.E., Seward F.D., et al., 1991, \apjs, 76, 1127

\noindent [262] Strohmayer T.E. \& Markwardt C.B., 2002, \apj, 577, 337

\noindent [263] Galloway D.K., Psaltis D., Muno M.P., \& Chakrabarty D., 2006, \apj, 639, 1033

\noindent [264] Corbet R., Barbier L., Barthelmy S., et al., 2006, ATel, 779, 1

\noindent [265] Smith D.M., 2004, ATel, 338, 1

\noindent [266] Kuiper L., Jonker P., Hermsen W., \& O'Brien K., 2005, ATel, 654, 1

\noindent [267] Markwardt C.B., Swank J.H., \& Smith E., 2005, ATel, 465, 1

\noindent [268] Ebisawa K., Bourban G., Bodaghee A., et al., 2003, \aap, 411, L59

\noindent [269] Manchester R.N., Lyne A.G., Camilo F., et al., 2001, \mnras, 328, 17

\noindent [270] Netzer H., Lemze D., Kaspi S., et al., 2005, \apj, 629, 739

\noindent [271] Donato D., Sambruna R.M., \& Gliozzi M., 2005, \aap, 433, 1163

\noindent [272] Brocksopp C., McGowan K.E., Krimm H., et al., 2006, \mnras, 365, 1203

\noindent [273] Tueller J., Markwardt C., Ajello M., et al., 2006, ATel, 835, 1

\noindent [274] Kinnunen T. \& Skiff B.A., 2000, IBVS, 4863, 1

\noindent [275] Oosterbroek T., Parmar A.N., Dal Fiume D., et al., 2000, \aap, 353, 575

\noindent [276] Kuster M., Wilms J., Staubert R., et al., 2005, \aap, 443, 753

\noindent [277] Reynolds A.P., Quaintrell H., Still M.D., et al., 1997, \mnras, 288, 43

\noindent [278] Sugizaki M., Mitsuda K., Kaneda H., et al., 2001, \apjs, 134, 77

\noindent [279] Chakrabarty D., Wang Z., Juett A.M., et al., 2002, \apj, 573, 789

\noindent [280] Orlandini M., dal Fiume D., del Sordo S., et al., 1999, \aap, 349, L9

\noindent [281] Baykal A., 2000, \mnras, 313, 637

\noindent [282] Chakrabarty D., Grunsfeld J.M., Prince T.A., et al., 1993, \apjl, 403, L33

\noindent [283] Audley M.D., Nagase F., Mitsuda K., et al., 2006, \mnras, 367, 1147

\noindent [284] Krauss M.I., Juett A.M., Chakrabarty D., et al., 2006, ATel, 777, 1

\noindent [285] Kennea J.A., Marshall F.E., Steeghs D., et al., 2006, ATel, 704, 1

\noindent [286] Miller J.M., Raymond J., Fabian A.C., et al., 2004, \apj, 601, 450

\noindent [287] Hynes R.I., Steeghs D., Casares J., et al., 2004, \apj, 609, 317

\noindent [288] Boroson B., Vrtilek S.D., Kallman T., \& Corcoran M., 2003, \apj, 592, 516

\noindent [289] Hong J. \& Hailey C.J., 2004, \apj, 600, 743

\noindent [290] Ankay A., Kaper L., de Bruijne J.H.J., et al., 2001, \aap, 370, 170

\noindent [291] Kazarovets E.V., Samus N.N., \& Durlevich O.V., 2000, IBVS, 4870, 1

\noindent [292] Iaria R., Di Salvo T., Robba N.R., et al., 2004, \apj, 600, 358

\noindent [293] Wachter S. \& Margon B., 1996, \aj, 112, 2684

\noindent [294] Oosterbroek T., Penninx W., van der Klis M., et al., 1991, \aap, 250, 389

\noindent [295] Asai K., Dotani T., Hoshi R., et al., 1998, \pasj, 50, 611

\noindent [296] Remillard R.A., Orosz J.A., McClintock J.E., \& Bailyn C.D., 1996, \apj, 459, 226

\noindent [297] Rea N., Oosterbroek T., Zane S., et al., 2005, \mnras, 361, 710

\noindent [298] in 't Zand J.J.M., Cornelisse R., \& M\'endez M., 2005, \aap, 440, 287

\noindent [299] Di Salvo T., Iaria R., M\'endez M., et al., 2005, \apjl, 623, L121

\noindent [300] in 't Zand J.J.M., Heise J., Lowes P., \& Ubertini P., 2003, ATel, 160, 1

\noindent [301] Negueruela I. \& Schurch M., 2007, \aap, 461, 631

\noindent [302] Jonker P.G., M\'endez M., Nelemans G., et al., 2003, \mnras, 341, 823

\noindent [303] Kennea J.A., Burrows D.N., Nousek J.A., et al., 2005, ATel, 476, 1

\noindent [304] Ritter H. \& Kolb U., 2003, \aap, 404, 301

\noindent [305] Wilson C.A., Patel S.K., Kouveliotou C., et al., 2003, \apj, 596, 1220

\noindent [306] Edge A.C. \& Stewart G.C., 1991, \mnras, 252, 414

\noindent [307] Nevalainen J., Oosterbroek T., Bonamente M., \& Colafrancesco S., 2004, \apj, 608, 166

\noindent [308] Cocchi M., Bazzano A., Natalucci L., et al., 2001, \memsai, 72, 757

\noindent [309] Downes R.A., Webbink R.F., Shara M.M., et al., 2001, \pasp, 113, 764

\noindent [310] de Martino D., Matt G., Belloni T., et al., 2004, \aap, 415, 1009

\noindent [311] Norton A.J., Wynn G.A., \& Somerscales R.V., 2004, \apj, 614, 349

\noindent [312] Suleimanov V., Revnivtsev M., \& Ritter H., 2005, \aap, 435, 191

\noindent [313] Cornelisse R., Charles P.A., \& Robertson C., 2006, \mnras, 366, 918

\noindent [314] Cadolle Bel M., Rodriguez J., Sizun P., et al., 2004, \aap, 426, 659

\noindent [315] Nagata T., Kato D., Baba D., et al., 2003, \pasj, 55, L73

\noindent [316] Bykov A.M., Krassilchtchikov A.M., Uvarov Y.A., et al., 2006, \aap, 449, 917

\noindent [317] Zurita Heras J.A., de Cesare G., Walter R., et al., 2006, \aap, 448, 261

\noindent [318] Corbet R.H.D., Markwardt C.B., \& Swank J.H., 2005, \apj, 633, 377

\noindent [319] Steeghs D., Torres M.A.P., Koviak K., et al., 2005, ATel, 629, 1

\noindent [320] Kennea J.A., Burrows D.N., Chester M., et al., 2005, ATel, 632, 1

\noindent [321] Kuulkers E., den Hartog P.R., in 't Zand J.J.M., et al., 2003, \aap, 399, 663

\noindent [322] G\"ansicke B.T., Marsh T.R., Edge A., et al., 2005, \mnras, 361, 141

\noindent [323] Church M.J. \& Baluci\'nska-Church M., 2001, \aap, 369, 915

\noindent [324] D'A\'i, A., di Salvo T., Iaria R., et al., 2006, \aap, 448, 817

\noindent [325] Strohmayer T.E., Zhang W., Swank J.H., et al., 1996, \apjl, 469, L9

\noindent [326] Hernanz M. \& Sala G., 2002, Science, 298, 393

\noindent [327] Munari U. \& Zwitter T., 2002, \aap, 383, 188

\noindent [328] Paul B., Dotani T., Nagase F., et al., 2005, \apj, 627, 915

\noindent [329] Hinkle K.H., Fekel F.C., Joyce R.R., et al., 2006, \apj, 641, 479

\noindent [330] Chelovekov I.V., Grebenev S.A., \& Sunyaev R.A., 2006, AstL, 32, 456

\noindent [331] Marti J., Mirabel I.F., Chaty S., \& Rodriguez L.F., 1998, \aap, 330, 72

\noindent [332] Molina M., Malizia A., Bassani L., et al., 2006, \mnras, 371, 821

\noindent [333] Corbet R.H.D., Thorstensen J.R., Charles P.A., et al., 1986, \mnras, 222, 15P

\noindent [334] Smith D.M., Heindl W.A., Markwardt C.B., et al., 2006, \apj, 638, 974

\noindent [335] Sakano M., Koyama K., Murakami H., et al., 2002, \apjs, 138, 19

\noindent [336] Negueruela I., Smith D.M., Harrison T.E., \& Torrej\'on J.M., 2006, \apj, 638, 982

\noindent [337] Torres M.A.P., Steeghs D., Garcia M.R., et al., 2006, ATel, 784, 1

\noindent [338] in 't Zand J.J.M., Verbunt F., Kuulkers E., et al., 2002, \aap, 389, L43

\noindent [339] Sidoli L., Belloni T., \& Mereghetti S., 2001, \aap, 368, 835

\noindent [340] Torres M.A.P., Garcia M.R., McClintock J.E., et al., 2004, ATel, 264, 1

\noindent [341] Kong A.K.H., 2006, ATel, 745, 1

\noindent [342] Kennea J.A., Burrows D.N., Nousek J., \& Gehrels N., 2005, ATel, 617, 1

\noindent [343] Rupen M.P., Dhawan V., \& Mioduszewski A.J., 2004, ATel, 335, 1

\noindent [344] Mart\'i, J., Mirabel I.F., Chaty S., \& Rodr\'iguez L.F., 2000, \aap, 363, 184

\noindent [345] Gallo E. \& Fender R.P., 2002, \mnras, 337, 869

\noindent [346] Smith D.M., Heindl W.A., \& Swank J.H., 2002, \apjl, 578, L129

\noindent [347] White N.E. \& van Paradijs J., 1996, \apjl, 473, L25

\noindent [348] Reid M.J. \& Brunthaler A., 2004, \apj, 616, 872

\noindent [349] B\'elanger G., Goldwurm A., Renaud M., et al., 2006, \apj, 636, 275

\noindent [350] Miller J.M., Raymond J., Homan J., et al., 2006, \apj, 646, 394

\noindent [351] Corbel S., Kaaret P., Fender R.P., et al., 2005, \apj, 632, 504

\noindent [352] Porquet D., Rodriguez J., Corbel S., et al., 2003, \aap, 406, 299

\noindent [353] Revnivtsev M.G., Sunyaev R.A., Varshalovich D.A., et al., 2004, AstL, 30, 382

\noindent [354] Wijnands R., Miller J.M., \& Wang Q.D., 2002, \apj, 579, 422

\noindent [355] Kong A.K.H., Wijnands R., \& Homan J., 2005, ATel, 641, 1

\noindent [356] Werner N., in 't Zand J.J.M., Natalucci L., et al., 2004, \aap, 416, 311

\noindent [357] Revnivtsev M.G., Churazov E.M., Sazonov S.Y., et al., 2004, \aap, 425, L49

\noindent [358] Juett A.M., Kaplan D.L., Chakrabarty D., et al., 2005, ATel, 521, 1

\noindent [359] Kennea J.A., Burrows D.N., Markwardt C., \& Gehrels N., 2005, ATel, 500, 1

\noindent [360] Oosterbroek T., Barret D., Guainazzi M., \& Ford E.C., 2001, \aap, 366, 138

\noindent [361] McClintock J., Murray S., Garcia M., \& Jonker P., 2003, ATel, 205, 1

\noindent [362] Bhattacharyya S., Strohmayer T.E., Markwardt C.B., \& Swank J.H., 2006, \apjl, 639, L31

\noindent [363] Torii K., Kinugasa K., Katayama K., et al., 1998, \apj, 508, 854

\noindent [364] Paizis A., Nowak M.A., Rodriguez J., et al., 2006, ATel, 907, 1

\noindent [365] Sidoli L., Parmar A.N., Oosterbroek T., et al., 2001, \aap, 368, 451

\noindent [366] in 't Zand J.J.M., Hulleman F., Markwardt C.B., et al., 2003, \aap, 406, 233

\noindent [367] Still M., 2005, ATel, 555, 1

\noindent [368] Morris D.C., Burrows D.N., Racusin J., et al., 2005, ATel, 552, 1

\noindent [369] T\"urler M., Shaw S.E., Kuulkers E., et al., 2006, ATel, 790, 1

\noindent [370] in 't Zand J.J.M., 2005, \aap, 441, L1

\noindent [371] Pellizza L.J., Chaty S., \& Negueruela I., 2006, \aap, 455, 653

\noindent [372] Lutovinov A., Revnivtsev M., Molkov S., \& Sunyaev R., 2005, \aap, 430, 997

\noindent [373] Sidoli L. \& Mereghetti S., 2002, \aap, 388, 293

\noindent [374] Iaria R., di Salvo T., Robba N.R., et al., 2005, \aap, 439, 575

\noindent [375] Hill A.B., Walter R., Knigge C., et al., 2005, \aap, 439, 255

\noindent [376] Combi J.A., Rib\'o M., Mart\'i J., \& Chaty S., 2006, \aap, 458, 761

\noindent [377] Markwardt C.B., Juda M., \& Swank J.H., 2003, \iaucirc, 8095, 2

\noindent [378] Falanga M., Bonnet-Bidaud J.M., Poutanen J., et al., 2005, \aap, 436, 647

\noindent [379] Kaplan D.L., Fox D.W., Kulkarni S.R., et al., 2002, \apj, 564, 935

\noindent [380] Molkov S., Hurley K., Sunyaev R., et al., 2005, \aap, 433, L13

\noindent [381] Tiengo A., Esposito P., Mereghetti S., et al., 2005, \aap, 440, L63

\noindent [382] Corbel S. \& Eikenberry S.S., 2004, \aap, 419, 191

\noindent [383] Kaspi V.M., Roberts M.E., Vasisht G., et al., 2001, \apj, 560, 371

\noindent [384] Roberts M.S.E., Tam C.R., Kaspi V.M., et al., 2003, \apj, 588, 992

\noindent [385] Torii K., Tsunemi H., Dotani T., et al., 1999, \apjl, 523, L69

\noindent [386] Gotthelf E.V., 2003, \apj, 591, 361

\noindent [387] Ubertini P., Bassani L., Malizia A., et al., 2005, \apjl, 629, L109

\noindent [388] Brogan C.L., Gaensler B.M., Gelfand J.D., et al., 2005, \apjl, 629, L105

\noindent [389] Ueda Y., Murakami H., Yamaoka K., et al., 2004, \apj, 609, 325

\noindent [390] Corbet R.H.D., 2003, \apj, 595, 1086

\noindent [391] Bandyopadhyay R.M., Shahbaz T., Charles P.A., \& Naylor T., 1999, \mnras, 306, 417

\noindent [392] Sguera V., Bird A.J., Dean A.J., et al., 2006, ATel, 873, 1

\noindent [393] Farinelli R., Frontera F., Zdziarski A.A., et al., 2005, \aap, 434, 25

\noindent [394] Dhawan M.P.R.V. \& Mioduszewski A.J., 2006, ATel, 721, 1

\noindent [395] Miller J.M., Homan J., Steeghs D., \& Wijnands R., 2006, ATel, 746, 1

\noindent [396] Sala G. \& Greiner J., 2006, ATel, 791, 1

\noindent [397] Gehrels M.S.N., Steeghs D., Torres M.A.P., et al., 2005, ATel, 588, 1

\noindent [398] in 't Zand J., Jonker P., Mendez M., \& Markwardt C., 2006, ATel, 915, 1

\noindent [399] Kinugasa K., Torii K., Hashimoto Y., et al., 1998, \apj, 495, 435

\noindent [400] Voges W., Aschenbach B., Boller T., et al., 2000, \iaucirc, 7432, 1

\noindent [401] Migliari S., Fender R.P., Rupen M., et al., 2004, \mnras, 351, 186

\noindent [402] Revnivtsev M., Sazonov S., Churazov E., \& Trudolyubov S., 2006, \aap, 448, L49

\noindent [403] Douglas J.N., Bash F.N., Bozyan F.A., et al., 1996, \aj, 111, 1945

\noindent [404] Juett A.M. \& Chakrabarty D., 2005, \apj, 627, 926

\noindent [405] Iaria R., Di Salvo T., Burderi L., \& Robba N.R., 2001, \apj, 557, 24

\noindent [406] Jonker P.G. \& van der Klis M., 2001, \apjl, 553, L43

\noindent [407] Hellier C. \& Mason K.O., 1989, \mnras, 239, 715

\noindent [408] Mason K.O. \& Cordova F.A., 1982, \apj, 262, 253

\noindent [409] Monet D. \& et al. 1998, in The PMM USNO-A2.0 Catalog. (1998), 0--+

\noindent [410] Martocchia A., Motch C., \& Negueruela I., 2005, \aap, 430, 245

\noindent [411] Casares J., Rib\'o M., Ribas I., et al., 2005, \mnras, 364, 899

\noindent [412] Thompson T.W.J., Rothschild R.E., Tomsick J.A., \& Marshall H.L., 2005, \apj, 634, 1261

\noindent [413] Homer L., Charles P.A., \& O'Donoghue D., 1998, \mnras, 298, 497

\noindent [414] Kong A.K.H., Homer L., Kuulkers E., et al., 2000, \mnras, 311, 405

\noindent [415] Safi-Harb S., Harrus I.M., Petre R., et al., 2001, \apj, 561, 308

\noindent [416] Camilo F., Ransom S.M., Gaensler B.M., et al., 2006, \apj, 637, 456

\noindent [417] Schmitz M., 1991, NED Team Report, 1, 1 (1991), 1, 1

\noindent [418] de Rosa A., Piro L., Tramacere A., et al., 2005, \aap, 438, 121

\noindent [419] Grandi P., Urry C.M., \& Maraschi L., 2002, \nar, 46, 221

\noindent [420] Parmar A.N., Oosterbroek T., Sidoli L., Stella L., \& Frontera F., 2001, \aap, 380, 490

\noindent [421] Deutsch E.W., Margon B., \& Anderson S.F., 2000, \apjl, 530, L21

\noindent [422] Malizia A., Bassani L., Stephen J.B., et al., 2005, \apjl, 630, L157

\noindent [423] Foster R.S., Cadwell B.J., Wolszczan A., \& Anderson S.B., 1995, \apj, 454, 826

\noindent [424] Molkov S.V., Cherepashchuk A.M., Lutovinov A.A., et al., 2004, AstL, 30, 534

\noindent [425] Halpern J.P., Gotthelf E.V., Helfand D.J., et al., 2004, ATel, 289, 1

\noindent [426] Bamba A., Yokogawa J., Ueno M., et al., 2001, \pasj, 53, 1179

\noindent [427] Wachter S., Patel S.K., Kouveliotou C., et al., 2004, \apj, 615, 887

\noindent [428] Morii M., Sato R., Kataoka J., \& Kawai N., 2003, \pasj, 55, L45

\noindent [429] Vasisht G. \& Gotthelf E.V., 1997, \apjl, 486, L129

\noindent [430] Sanbonmatsu K.Y. \& Helfand D.J., 1992, \aj, 104, 2189

\noindent [431] Sguera V., Bird A.J., Dean A.J., et al., 2007, \aap, 462, 695

\noindent [432] Coe M.J., Fabregat J., Negueruela I., et al., 1996, \mnras, 281, 333

\noindent [433] Piraino S., Santangelo A., Segreto A., et al., 2000, \aap, 357, 501

\noindent [434] Israel G.L., Negueruela I., Campana S., et al., 2001, \aap, 371, 1018

\noindent [435] Gotthelf E.V., Vasisht G., Boylan-Kolchin M., \& Torii K., 2000, \apjl, 542, L37

\noindent [436] Helfand D.J., Collins B.F., \& Gotthelf E.V., 2003, \apj, 582, 783

\noindent [437] Livingstone M.A., Kaspi V.M., Gotthelf E.V., \& Kuiper L., 2006, \apj, 647, 1286

\noindent [438] Becker R.H. \& Helfand D.J., 1984, \apj, 283, 154

\noindent [439] Soffitta P., Tomsick J.A., Harmon B.A., et al., 1998, \apjl, 494, L203

\noindent [440] Lutovinov A.A. \& Revnivtsev M.G., 2003, AstL, 29, 719

\noindent [441] Beuermann K., Harrison T.E., McArthur B.E., et al., 2004, \aap, 419, 291

\noindent [442] Corbet R.H.D. \& Mukai K., 2002, \apj, 577, 923

\noindent [443] Corbet R.H.D., Marshall F.E., Peele A.G., \& Takeshima T., 1999, \apj, 517, 956

\noindent [444] Harris D.E., Forman W., Gioa I.M., et al., 1996, VizieR Online Data Catalog, 9013, 0

\noindent [445] Bikmaev I.F., Sunyaev R.A., Revnivtsev M.G., \& Burenin R.A., 2005, AstL, in press, astro-ph/0511405

\noindent [446] Paul B. \& Rao A.R., 1998, \aap, 337, 815

\noindent [447] Takeshima T., Corbet R.H.D., Marshall F.E., et al., 1998, \iaucirc, 6826, 1

\noindent [448] Rupen M.P., Mioduszewski A.J., \& Dhawan V., 2005, ATel, 530, 1

\noindent [449] Campana S., 2005, ATel, 535, 1

\noindent [450] Kaaret P., Morgan E.H., Vanderspek R., \& Tomsick J.A., 2006, \apj, 638, 963

\noindent [451] Kawai N. \& Suzuki M., 2005, ATel, 534, 1

\noindent [452] Galloway D.K., Remillard R., \& Morgan E., 2003, \iaucirc, 8081, 2

\noindent [453] Galloway D.K., Wang Z., \& Morgan E.H., 2005, \apj, 635, 1217

\noindent [454] Frail D.A., Kulkarni S.R., \& Bloom J.S., 1999, \nat, 398, 127

\noindent [455] Esposito P., Mereghetti S., Tiengo A., et al., 2007, \aap, 461, 605

\noindent [456] Vrba F.J., Henden A.A., Luginbuhl C.B., et al., 2000, \apjl, 533, L17

\noindent [457] Rupen M.P., Dhawan V., \& Mioduszewski A.J., 2002, \iaucirc, 7874, 1

\noindent [458] in 't Zand J.J.M., Miller J.M., Oosterbroek T., \& Parmar A.N., 2002, \aap, 394, 553

\noindent [459] Chaty S., Mignani R.P., \& Israel G.L., 2006, \mnras, 365, 1387

\noindent [460] Cusumano G., di Salvo T., Burderi L., et al., 1998, \aap, 338, L79

\noindent [461] Baykal A., \.Inam S.C., \& Beklen E., 2006, \mnras, 369, 1760

\noindent [462] in 't Zand J.J.M., Baykal A., \& Strohmayer T.E., 1998, \apj, 496, 386

\noindent [463] Cox N.L.J., Kaper L., \& Mokiem M.R., 2005, \aap, 436, 661

\noindent [464] Levine A.M., Rappaport S., Remillard R., \& Savcheva A., 2004, \apj, 617, 1284

\noindent [465] Morel T. \& Grosdidier Y., 2005, \mnras, 356, 665

\noindent [466] Wijnands R., Maitra D., Bailyn C., \& Linares M., 2006, ATel, 871, 1

\noindent [467] Chevalier C. \& Ilovaisky S.A., 1991, \aap, 251, L11

\noindent [468] Brinkmann W., Kotani T., \& Kawai N., 2005, \aap, 431, 575

\noindent [469] Blundell K.M. \& Bowler M.G., 2004, \apjl, 616, L159

\noindent [470] in 't Zand J.J.M., Jonker P.G., Nelemans G., et al., 2006, \aap, 448, 1101

\noindent [471] Corbet R.H.D., Hannikainen D.C., \& Remillard R., 2004, ATel, 269, 1

\noindent [472] Martocchia A., Matt G., Belloni T., et al., 2006, \aap, 448, 677

\noindent [473] Harlaftis E.T. \& Greiner J., 2004, \aap, 414, L13

\noindent [474] Juett A.M. \& Chakrabarty D., 2006, \apj, 646, 493

\noindent [475] Falanga M., Belloni T., \& Campana S., 2006, \aap, 456, L5

\noindent [476] Sambruna R.M., 1997, \apj, 487, 536

\noindent [477] V\'eron-Cetty M.-P. \& V\'eron P., 2003, \aap, 412, 399

\noindent [478] Wang J., Wei J.Y., \& He X.T., 2006, \apj, 638, 106

\noindent [479] Andronov I.L., Baklanov A.V., \& Burwitz V., 2006, \aap, 452, 941

\noindent [480] Rana V.R., Singh K.P., Barrett P.E., \& Buckley D.A.H., 2005, \apj, 625, 351

\noindent [481] Naik S., Callanan P.J., Paul B., \& Dotani T., 2006, \apj, 647, 1293

\noindent [482] Galloway D.K., Morgan E.H., \& Levine A.M., 2004, \apj, 613, 1164

\noindent [483] Tsygankov S.S. \& Lutovinov A.A., 2005, AstL, 31, 88

\noindent [484] Evans D.A., Worrall D.M., Hardcastle M.J., et al., 2006, \apj, 642, 96

\noindent [485] Liu F.K. \& Zhang Y.H., 2002, \aap, 381, 757

\noindent [486] Masetti N., Orlandini M., Palazzi E., et al., 2006, \aap, 453, 295

\noindent [487] Mattana F., G\"otz D., Falanga M., et al., 2006, \aap, 460, L1

\noindent [488] Schulz N.S., Cui W., Canizares C.R., et al., 2002, \apj, 565, 1141

\noindent [489] Zi\'olkowski J., 2005, \mnras, 358, 851

\noindent [490] Zhou H.-Y. \& Wang T.-G., 2002, \cjaa, 2, 501

\noindent [491] NED Team, 1992, NED Team Report, 1, 1 (1992), 1, 1

\noindent [492] Bykov A.M., Krassilchtchikov A.M., Uvarov Y.A., et al., 2006, \apjl, 649, L21

\noindent [493] Reig P. \& Coe M.J., 1999, \mnras, 302, 700

\noindent [494] Camero Arranz A., Wilson C.A., Connell P., et al., 2005, \aap, 441, 261

\noindent [495] Wilson C.A., Finger M.H., Coe M.J., et al., 2002, \apj, 570, 287

\noindent [496] Weekes T.C. \& Geary J.C., 1982, \pasp, 94, 708

\noindent [497] Stark M.J. \& Saia M., 2003, \apjl, 587, L101

\noindent [498] Lara L., Cotton W.D., Feretti L., et al., 2001, \aap, 370, 409

\noindent [499] Ballantyne D.R., 2005, \mnras, 362, 1183

\noindent [500] Cutri R.M., Skrutskie M.F., van Dyk S., et al., 2003, The IRSA 2MASS All-Sky Point
  Source Catalog, NASA/IPAC Infrared Science Archive. (http://irsa.ipac.caltech.edu/applications/Gator/)

\noindent [501] Baykal A., Stark M.J., \& Swank J.H., 2002, \apj, 569, 903

\noindent [502] Sidoli L., Mereghetti S., Larsson S., et al., 2005, \aap, 440, 1033

\noindent [503] Falcone A., Tueller J., Markwardt C., et al., 2006, ATel, 846, 1

\noindent [504] Thorstensen J.R. \& Taylor C.J., 2001, \mnras, 326, 1235

\noindent [505] Halpern J.P., 2006, ATel, 847, 1

\noindent [506] Bonnet-Bidaud J.M., Mouchet M., de Martino D., et al., 2006, \aap, 445, 1037

\noindent [507] Motch C., Guillout P., Haberl F., et al., 1997, \aaps, 122, 201

\noindent [508] Mauche C.W., 2004, \apj, 610, 422

\noindent [509] Harrison T.E., McNamara B.J., Szkody P., \& Gilliland R.L., 2000, \aj, 120, 2649

\noindent [510] Costantini E., Freyberg M.J., \& Predehl P., 2005, \aap, 444, 187

\noindent [511] Casares J., Charles P.A., \& Kuulkers E., 1998, \apjl, 493, L39

\noindent [512] Stickel M., Lemke D., Klaas U., et al., 2004, \aap, 422, 39

\noindent [513] Dadina M., Bassani L., Cappi M., et al., 2001, \aap, 370, 70

\noindent [514] Ravasio M., Tagliaferri G., Ghisellini G., et al., 2003, \aap, 408, 479

\noindent [515] Hog E., Kuzmin A., Bastian U., et al., 1998, \aap, 335, L65

\noindent [516] Masetti N., Dal Fiume D., Amati L., et al., 2004, \aap, 423, 311

\noindent [517] Rib\'o M., Negueruela I., Blay P., et al., 2006, \aap, 449, 687

\noindent [518] Blay P., Negueruela I., Reig P., et al., 2006, \aap, 446, 1095

\noindent [519] Patterson J., Kemp J., Richman H.R., et al., 1998, \pasp, 110, 415

\noindent [520] Barrett P., 1996, \pasp, 108, 412

\noindent [521] Gibson R.R., Marshall H.L., Canizares C.R., \& Lee J.C., 2005, \apj, 627, 83

\noindent [522] Scott J.E., Kriss G.A., Lee J.C., et al., 2005, \apj, 634, 193

\noindent [523] George I.M., Turner T.J., Netzer H., et al., 1998, \apjs, 114, 73

\noindent [524] Hwang U., Laming J.M., Badenes C., et al., 2004, \apjl, 615, L117

\noindent [525] Reed J.E., Hester J.J., Fabian A.C., \& Winkler P.F., 1995, \apj, 440, 706

\clearpage
\twocolumn

\end{document}